\documentclass[aps,preprint]{revtex4}
\def\be{\begin{equation}}
\def\ee{\end{equation}}
\def\bear{\begin{eqnarray}}
\def\eear{\end{eqnarray}}
\def\bearst{\begin{eqnarray*}}
\def\eearst{\end{eqnarray*}}
\usepackage[latin1]{inputenc}
\usepackage[english]{babel}
\usepackage[T1]{fontenc}
\usepackage[dvips]{graphicx}
\usepackage{amssymb}
\usepackage{epsfig,bbm}
\newtheorem{teorema}{Theorem} 
\begin{document}

\title{Free-fall in a uniform gravitational field in non-commutative quantum mechanics}

\author{K. H. C. Castello-Branco\footnote{Present address: 
Universidade Federal do Oeste do Pará, Av. Marechal Rondon, s/n, Santar\'em, Pará, 68040-070, Brazil}}
\email{khccb@yahoo.com.br}
\affiliation{Instituto de F\'isica de S\~ao Carlos, Universidade de S\~ao Paulo, Av. Trabalhador S\~ao-Carlense, 400, S\~ao Carlos,
S\~ao Paulo 13560-970, Brazil}

\author{A. G. Martins}
\email{andrey_martins@yahoo.com.br}
\affiliation{Departamento de Ci\^encias Naturais, Universidade do Estado do Par\'a, Av. Djalma Dutra, s/n, Bel\'em, Pará, 66113-200, Brazil.}
\begin{abstract}


We study the free-fall of a quantum particle in the context of
Non-Commutative Quantum Mechanics (NCQM). 
Assuming non-commuta\-tivity of 
the canonical type between the coordinates of a two dimensional 
configuration space, we consider a neutral particle trapped in a
gravitational well and exactly solve the energy eigenvalue problem.
By resorting to experimental data from the GRANIT experiment, in
which the first energy levels of freely falling quantum ultracold 
neutrons were determined, we impose an upper-bound on the non-commutativity 
parameter. We also investigate the time of flight of a quantum 
particle moving in a uniform gravitational field in NCQM. This is 
related to the Weak Equivalence Principle. As we consider stationary, energy 
eigenstates, i.e., delocalized states, the time of flight must be 
measured by a quantum clock, suitably coupled to the particle. 
By considering the clock as a small perturbation, we solve the (stationary) 
scattering problem associated and show that the time of flight is equal to the  
classical result, when the measurement is made far from the 
turning point. This result is interpreted as an extension of the 
Equivalence Principle to the realm of NCQM.

\end{abstract}



\maketitle


\section{Introduction}\label{introduction-paper}

The idea that space-time would have a non-commutative
structure was proposed by Heisenberg and others 
\cite{Meyenn:1985}, already at the very beginning of Quantum
Field Theory (QFT), so that an effective ultraviolet cutoff could be
introduced at very small length scales, in an attempt to get rid of 
divergences. It was only in 1947, with Snyder \cite{Snyder:1946qz}, 
that this idea of space-time non-commutativity was formalized 
(for a historical introduction, see \cite{Szabo:2001kg,Jackiw:2001dj}). But due to  
the success achieved by the renormalization program of QFTs,  
relatively low interest remained on the subject. In 
the late 1990s some results coming from String Theory have 
suggested that space-time may display a non-commutative structure 
\cite{Connes:1997cr}, thus starting a great revival on the study of 
QFTs based on non-commuting space-time coordinates (for reviews, 
see \cite{Szabo:2001kg,Douglas:2001ba,Girotti:2003at}). Nevertheless, we 
remark that the issue of space-time non-commutativity was considered 
in an earlier work by Doplicher et al., who constructed a unitary QFT 
based on a non-commutative space-time \cite{Doplicher:1994zv}.
motivated by the issue of the description of the quantum nature of the 
space-time.
 
As quantum mechanics can be considered the one-particle
sector of quantum field theory, it is interesting to
study 
the quantum mechanics defined on non-commutative spaces. Lately,
this Non-Commutative Quantum Mechanics (NCQM) has been
increasingly studied (see, for instance, Ref.s \cite{Gamboa:2001yq,Mezincescu:2000zq,Nair:2000ii}) and
effects of non-commutativity that might be experimentally
detectable have been investigated \cite{Wu:2006jx,Zhang:2006sg}. 
There have been also studies of many-particle systems
\cite{Luo:2005je} and an approach to NCQM that is directly 
related to the ideas developed in \cite{Doplicher:1994zv} has been developed in \cite{Balachandran:2004rq}, 
by means of an algebraic setup in which the time is a non-commutative 
coordinate as well as the space coordinates, with interesting 
ensuing consequences (see Ref. \cite{Balachandran:2004yh}).

In this paper we are concerned about the physics of a quantum particle
in a uniform gravitational field. In the first part of the paper we study the problem known as the {\it gravitational quantum well}.
This can be obtained by the Earth's gravitational field and
a perfectly reflecting mirror at the bottom. In the context of ordinary quantum mechanics 
this well-known problem has been
thoroughly studied in text-books and
pedagogic articles (see, for instance, 
\cite{Langhoff:1971pt}). But this simple theoretical 
investigation gained a lot in importance, as recently such a quantum 
well was experimentally realized and the first two quantum 
states of neutrons moving in the Earth's gravitational well were 
identified. This was achieved through the GRANIT experiment,
 performed by Nesvizhevsky et al. \cite{Nesvizhevsky:2002,Nesvizhevsky:2005ss}.
In this paper we study the non-commutative
version of this problem, by means of the solution of the 
non-commutative Schr\"odinger equation. As it will be shown, the 
experimental results obtained in \cite{Nesvizhevsky:2002,Nesvizhevsky:2005ss} 
allow us to find an upper-bound on the parameter
of the spatial non-commutativity. The first study to treat the 
non-commutative gravitational quantum well was done by Bertolami 
et al. \cite{Bertolami:2005jw}, but in a non-commutative 
model different from the one we will consider in this paper. 
We remark that we have assumed only spatial non-commutativity, in contrast to other works 
in the literature, which considered non-commutativity of both configuration
and momentum spaces (see \cite{Bertolami:2005jw}-\cite{Kochan:2001pz}) or time-space non-commutativity (see \cite{Saha:2007}). 
The Ref.s \cite{Bertolami:2005jw}-\cite{Buisseret:2007qd} treated the non-commutative gravitational quantum well and used data from
the GRANIT experiment \cite{Nesvizhevsky:2002,Nesvizhevsky:2005ss}
to find upper-bounds on the value of the {\it momentum-momentum}
non-commutativity parameter, while in the Ref. \cite{Saha:2007} an upper bound on the {\it time-space} component of the
non-commutative matrix was found, by means of second quantization techniques.
Nonetheless, differently from \cite{Bertolami:2005jw}-\cite{Buisseret:2007qd} and \cite{Saha:2007},
we have found the shifts in the energy levels
due to spatial non-commutativity only. This was done by studying the adjointness of the Hamiltonian operator 
associated with the problem, what led us  to determine the self-adjoint extensions of it.  We have treated the problem 
without resorting to any perturbative approach.

Although studies of non-commutative corrections to general relativity have been addressed 
in the literature (see, e.g., \cite{Harikumar:2006xf}), in our work we study the quantum mechanics of 
a particle with non-commutative spatial degrees of freedom and subject to a Newtonian 
gravitational field. In fact, since in the GRANIT experiment one considers the gravitational 
field near the Earth's surface, where the Newtonian gravity is an excellent approximation, 
 one can treat quantum mechanics with spatial non-commutative degrees of freedom 
as a fairly well approximation, without evoking the possible non-commutative structure that gravity might display. 
As we are not considering gravity as being described by general relativity, we did not care about 
non-commutative corrections to the Hilbert-Einstein action. 
Furthermore, the spatial non-commuting coordinates that we are considering are not the coordinates of the underlying 
space-time, but rather of the configuration space. 
The assumption of non-commutativity of the configuration 
space does not necessarily imply a non-commutative space-time. One 
can, for instance, study a non-commutative scalar field coupled to gravity in
a curved space-time, without taking into account modifications of gravity due
to non-commutativity (see, e.g., \cite{Bertolami:2002eq}). We stress that we are studying 
a non-relativistic quantum mechanics and that the non-commutativity in this case is
conceptually different from that of relativistic quantum mechanics or quantum 
field theory, where the space-time itself is to be considered non-commutative.
As remarked, for instance, in the Ref.s \cite{Girotti:2004,Bemfica:2008zza}, in NCQM the
non-commutativity modifies only the algebra of the basic observables of the
theory and does not introduce modifications in the structure of the underlying
space-time (see also \cite{Barbosa:2003cq} and references therein).

In the second part of this paper we consider the issue of the Equivalence Principle
in the context of NCQM. In ordinary quantum mechanics, this is an interesting and
subtle question (see, e.g., \cite{Lammerzahl:1996se}).
We address the issue of the (Weak) Equivalence Principle in NCQM motivated by the study due to 
Davies in the case of ordinary quantum mechanics \cite{Davies:2006}.
Although (\emph{i}$\,$) classically the Equivalence Principle has a {\it local} 
character, and (\emph{ii}$\,$) in a uniform gravitational field the 
Schr\"odinger equation leads to mass-dependent results, one can 
still say that the Equivalence Principle is realized at 
the quantum level if the motion of a quantum particle is considered to 
be described by a wave packet. This is a consequence of 
the Ehrenfest's theorem (see discussion in \cite{Davies:2006}). 
Nevertheless, Davies asks about the validity of the Equivalence Principle  
in the case of \emph{delocalized} quantum states, such as energy eigenstates. 
These do not have classical counterparts associated to a localized particle with 
a well-defined trajectory. In order to analyse this, Davies 
considered a variant of the ({\it gedanken}) experiment of Galileo at the 
leaning Tower of Pisa, with particles of different masses that would be 
vertically projected up in a uniform gravitational field. 
For classical particles, in the neighbourhood of the 
Earth, for instance, the out-and-back time is twice that spent in climbing, 
but for quantum particles there is a non-vanishing probability 
that they tunnel into the classically forbidden region. 
This special feature of the quantum theory might then give rise 
to a departure from the classical turnaround time, since a delay 
might ensue. In ordinary quantum mechanics, when 
one considers stationary, energy eigenstates, 
Davies showed that, when the measurement 
is made far from the classical turning point, the time of 
flight of a quantum particle is identical to the classical result \cite{Davies:2006}. 
In this sense the (Weak) Equivalence Principle is preserved in quantum 
mechanics. In order to measure the time of flight associated 
with delocalized states, Davies applied a simple model 
of quantum clock, due to Peres \cite{Peres:1980}. 
This clock runs only when the particle travels within a given region 
of interest and does not measure absolute instants of time, but just the 
time difference.  

It is in the sense mentioned above that in this paper we show that the Equivalence 
Principle can be extended to the NCQM in the case of stationary, energy eigenstates. Nevertheless, 
we remark that we achieve this result by explicitly solving the (stationary) 
scattering problem of a quantum particle in the presence of a uniform gravitational 
field. Our approach to the problem does not closely follow that of Davies, but rather we 
closely follow the original approach by Peres \cite{Peres:1980} 
and explicitly consider that the quantum clock is coupled to the particle. 
Then, we will show that, in the case of small interaction between 
the particle and the clock, and when the measurement is made far from the 
turning point, the time of flight will be given by the \emph{phase shift} 
of the wave function and this shift correctly codifies the 
time as measured by the quantum clock. 

This paper is organized as follows: in Sec. \ref{quantum-mech-non-commut}, we review the basic features of the formalism
of NCQM; in Sec. \ref{noncomm_quant_well}, we study the non-commutative gravitational 
quantum well; in Sec. \ref{granit-section}, we establish an upper-bound on the non-commutativity 
parameter; in Sec. \ref{time_section}, we review the Peres quantum clock model; in Sec. \ref{sec_tempo},
we apply it to the investigation of the time of flight of a particle subjected  
to a uniform gravitational field in NCQM, and in Sec. \ref{conclusao} we make concluding remarks.

\section{Non-commutative quantum mechanics}\label{quantum-mech-non-commut}

Quantum mechanics inspired the idea of non-commutative coordinates, which can be introduced by (see, for 
instance, \cite{Szabo:2001kg})
\be
[{\hat x}_{\mu},{\hat x}_{\nu}]=i\theta_{\mu\nu}\mathbbm{1}\,,
\label{space_time-nc_relation}
\ee
where $\,{\hat x}_{\mu}\,$ are the {\it coordinate operators} (the non-commutative
 analogues of the ordinary coordinate functions), $\,\mathbbm{1}\,$ is the unit element of the 
 non-commutative algebra and $\,\theta_{\mu\nu}\,$ 
is the anti-symmetric, real-valued $\,d\times d\,$ matrix describing the coordinate non-commutativity ($d\,$ is the space-time dimension). 
The uncertainty relations 
$\,\Delta x_\mu\Delta x_\nu \ge |\theta_{\mu\nu}| /2\,$, which are compatible with the commutation relations (\ref{space_time-nc_relation})
allows one to think of $\,\sqrt{|\theta_{\mu\nu}|}\,$ as the characteristic scale of 
length involved in the uncertainty
on the simultaneous measurement of the coordinates $\,\hat{x}_{\mu}\,$ and $\,\hat{x}_{\nu}\,$. In this paper we will set $\,\theta_{0\mu}=0\,$, since we 
are not interested in dealing with the possibility of violation of causality \cite{Seiberg:2000} and unitarity \cite{Gomis:2000}. 
The three-dimensional 
non-commutative space generated by $\hat{x}_{i}$ and $\mathbbm{1}$ will be denoted by $\,\mathbb{R}^{3}_{\theta}\,$.

In order to implement quantum physics we need to introduce the observables and the evolution equation. 
The position observables can be introduced by means of the left-action of $\,\mathbb{R}_{\theta}^{3}\,$ on itself, while
the momenta $\,\hat{p}_{i}\,$ commute with each other and are canonically conjugated to the position operators. 
The phase-space commutation relations then read
\be \label{esp-fases}
[{\hat x}_{i},{\hat x}_{j}]=i\theta_{ij}\,,\qquad[{\hat
 p}_{i},{\hat p}_{j}]=0\,,\qquad[{\hat p}_{i},{\hat
 x}_{j}]=-i\hbar\delta_{ij}\,. 
\ee
We remark that our phase-space commutation relations are distinct 
from the ones used in \cite{Bertolami:2005jw}-\cite{Buisseret:2007qd}, where not only the coordinates, but also the momenta 
are non-commutative. According to 
\cite{Bertolami:2005jw}, such a phase-space non-commutativity leads
to a modification of the Planck's constant, whereas 
\cite{Banerjee:2006jb} argues that this is not needed. Nevertheless, 
according to \cite{Bertolami:2005ud}, those approaches are in 
fact physically equivalent, differing only in the manner one defines 
the non-commutative parameters. On the other hand, according to \cite{Bertolami:2005jw,Govaerts:2007hi},
the type of phase-space algebra studied in \cite{Bertolami:2005jw}-\cite{Buisseret:2007qd} can be led to the 
canonical form by means of an appropriate change of variables. 
Moreover, this change of variables is not unique, as showed in \cite{Bastos:2006ps}.

The non-commutative phase-space operators in Eq. (\ref{esp-fases}) can be written in terms of the ordinary position and momentum
operators by means of the following (non-canonical) 
transformation (see \cite{Girotti:2004}, \cite{Bemfica:2008zza} and \cite{Kijanka:2004gs}, for example),
\be \label{equiv-algebras}
q_{i}=\hat{x}_{i}+\frac{\theta_{ij}}{2\hbar}\hat{p}_{j}\,, \qquad
p_{i}=\hat{p}_{i}\,,
\ee
where $\,q_i\,$ and $\,p_i\,$ satisfy $\,[q_i , p_j]=i\hbar\delta_{ij}\,$. 
Besides that,
we can represent the algebra (\ref{esp-fases}) of observables of NCQM
on the same representation space as the Heisenberg algebra (see, e.g., \cite{Girotti:2004,Bemfica:2008zza}), what means that
the set of states in NCQM is the same as in ordinary quantum mechanics.
This is interpreted as a manifestation of the fact that the non-commutativity introduced by 
Eq. (\ref{space_time-nc_relation}) has no observable consequences at 
the level of kinematics \cite{Govaerts:2007hi}. The non-trivial effects coming from the space-space non-commutativity 
are due to dynamical considerations.

Before we state the basic dynamical equation, we note that if we had considered a charged particle in an electromagnetic field, 
for instance, it would be quite natural to approach the problem of the dynamics via the Seiberg-Witten map (see, e.g., \cite{Kokado:2004um}).
But since the problem of a quantum neutral particle moving under the action of a Newtonian gravitational potential
is not related to a gauge theory, we cannot apply the gauge principle to study that interaction.
Hence, we will follow a quite standard approach that has been extensively
used in the literature on NCQM, which consists in starting with the noncommutative Schr\"odinger equation
%
\be
-\frac{\hbar ^{2}}{2m}\bigtriangledown ^{2}\Psi + V\star\Psi = i\hbar
\frac{\partial\Psi}{\partial t}\,,
\label{eq-schrodinger-NC}
\ee
where $\star$ denotes the so-called Moyal product, 
defined by 
\begin{eqnarray} \label{Moyal_product_definition}
(\psi\star\phi)(x_1,x_2,x_{3})=\psi(x_{1},x_{2},x_{3})\,
e^{\frac{i}{2}\overleftarrow{\partial}_{\!\! i}\,\,\theta_{ij}\overrightarrow{\partial}_{\! \! j}}\,
\phi(x_{1},x_{2},x_{3})\,,
\end{eqnarray}
which plays a major role, as it gives the action of the interaction Hamiltonian on wave functions.
We would like to emphazise that this approach has been applied by several authors to a variety of problems 
(see, for example, \cite{Gamboa:2001yq,Mezincescu:2000zq,Girotti:2004,Bemfica:2008zza}). 

An equivalent way to introduce the evolution equation (see, for example,
 \cite{Gamboa:2001yq}, \cite{Girotti:2004} and \cite{Bemfica:2008zza}) is to start with the ordinary Schr\"{o}dinger equation and then
substitute the ordinary (commutative) coordinates by the non-commutative ones, by making use of the inverse of Eq. (\ref{equiv-algebras}). 
As a consequence the interaction Hamiltonian gets modified by a shift (sometimes called \emph{Bopp shift}), that is,
\begin{eqnarray}\label{pot-moyal}
-\frac{\hbar ^{2}}{2m}\bigtriangledown ^{2}\Psi + 
V\left(x_{i}+\frac{i\theta_{ij}}{2}\frac{\partial}{\partial{x_{j}}}\right)\Psi=i\hbar\frac{\partial\Psi}{\partial t}\,.
\end{eqnarray}
By making use of the (\ref{Moyal_product_definition}) one can show that Eq. (\ref{pot-moyal})
is equivalent to Eq. (\ref{eq-schrodinger-NC}) (see, e.g., \cite{Girotti:2004}).

The energy eigenvalue equation of NCQM can be obtained by means of the usual stationary problem {\it ansatz}, $\,\Psi(x,y,z,t) =
 \psi(x,y,z)\,e^{-\frac{i}{\hbar}Et}\,$. The corresponding time-independent equation thus reads
\be
-\frac{\hbar ^{2}}{2m}\bigtriangledown ^{2}\psi + V\star\psi = 
E\psi\,,
\label{eq-schrodinger-NC-indep-t}
\ee
where $\,E\,$ is the energy of the particle.
 
Equations (\ref{eq-schrodinger-NC}) and (\ref{eq-schrodinger-NC-indep-t}) are the starting point of many calculations in NCQM 
that have been considered in the literature. They can be viewed as the Schr\"odinger 
equation of ordinary quantum  mechanics, but with the interaction term modified by a  
contribution coming from the non-commutativity of the configuration space. 
That means that the problems in NCQM can be approached 
in the same way as we would do in ordinary quantum mechanics, the only difference being the replacement of the ordinary 
potential energy by its deformed ($\theta_{ij}$-dependent) version. 
In this sense, the NCQM corresponds to a true deformation of ordinary quantum mechanics, and as
$\,\theta_{ij}\,$ goes to zero we recover the ordinary results. Some phenomenological consequences of this deformation will be addressed below, 
when we consider the two-dimensional motion of a quantum neutral particle at the presence of an external gravitational potential.

\section{The non-commutative gravitational quantum well}\label{noncomm_quant_well}

In this section we consider the non-commutative version of the gravitational 
quantum well and determine the energy spectrum of a particle trapped in it. 

Usually, the correct definition of an operator is not addressed in most applications 
in physics, but rather an operator is defined by means of its law of action (the so-called 
formal operator) only, with no mention about its domain of definition. Operators having the same formal 
expression but acting in different domains can lead to different physics. This is an 
important question specially in quantum theory \cite{Reed:1980}. Bearing this in mind, in order
to determine the spectrum of the non-commutative gravitational well, we will 
carefully examine the domain of the differential operator we have to handle with. 
This is closed related to the self-adjointness of the operator and will ultimately 
lead us to consider its self-adjoint extensions 
(an operator $\,S\,$ is an extension of an operator $\,T\,$
if $\,D(T)\subset D(S)\,$ and $\,S\phi=T\phi\,$, for all $\,\phi\in D(T)\,$, where $\,D(T)\,$ and $\,D(S)\,$ are the 
domains of definition of $\,T\,$ and $\,S\,$, respectively). This is a crucial step to correctly solve 
the eigenvalue problem associated to the quantum well. 

\subsection{The specification of the model}


Let $\,\vec{g}\,$ be a uniform gravitational field and let 
the $Oy$-axis be oriented in the opposite direction of 
$\,\vec{g}\,$, i.e., $\,\vec{g}=-g\,\hat{e}_{y}\,$.
At the boundary $\,y=0\,$ the particle encounters a perfectly reflecting mirror, which prevents it to 
get into the negative portion of the $Oy$ axis. We thus have 
\be
V(x,y,z)=
\left\{
\begin{array}{ccc}
mgy\,,    & \quad y > 0\,,\quad  & \forall x\,, z\,,\\
\infty\,, & \quad y=0\,,\quad  & \forall x\,, z\,.
\end{array}
\right. 
\label{potencial}
\ee

Since the particle configuration space is 
$\,Q=\mathbb{R}^{2}\times (0,\infty)\equiv \mathbb{R}^{2}\times\mathbb{R}_{+}^{*}\,$, we have to be cautious, 
because the Moyal product is not well-defined on manifolds with boundaries \cite{Balachandran:2003vm}, a fact that can be traced back to 
the non-locality of Eq. (\ref{Moyal_product_definition}). In order to avoid this problem we consider the $\,\theta_{ij}\,$ as 
sufficiently small parameters, so
that we can truncate the series of the Moyal product at some suitable power of $\,\theta_{ij}\,$. For example, the first 
order $\,\theta_{ij}\,$-dependent correction to the ordinary potential is
\begin{eqnarray}
V\star\psi  \simeq  V\,\psi 
+ \frac{i}{2}\theta_{ij}\frac{\partial V}{\partial x_{i}}
\frac{\partial}{\partial x_{j}}\psi\,.
\label{approx}
\end{eqnarray}
The most important feature of this approximation is its local character, which allows one to work in the realm of the upper half-space,
without having to worry about the issues coming from the non-locality of the Moyal product.
Thus, Eq. (\ref{approx}) leads to the following time-independent Schr\"{o}dinger equation,
\be
H\psi=-\frac{\hbar ^{2}}{2m}\nabla^{2}\psi
+ mgy\psi + \frac{img\theta_{21}}{2}\frac{\partial\psi}{\partial x}
+\frac{img\theta_{23}}{2}\frac{\partial\psi}{\partial z}=
E\psi \,.
\label{schrod-NC-cartes2}
\ee

It is interesting to notice that the direct application of  Eq. (\ref{Moyal_product_definition}) to the case of the 
potential in Eq. (\ref{potencial}) would lead to Eq. (\ref{schrod-NC-cartes2}) too: in the case of a simple linear potential,
the first order approximation considered in Eq. (\ref{approx}) is, in fact, exact. 
Nevertheless, in the presence of boundaries the Moyal product would lead to 
inconsistencies when applied to the calculation of non-linear quantities, such as transition amplitudes, provided that the wave function is non-polynomial.
Thus we will consider the approximation (\ref{approx}), as we said.

Although the Hamiltonian in Eq. (\ref{schrod-NC-cartes2}) acts on wave functions of the form $\,\psi(x,y,z)\,$, the 
non-commutative gravitational quantum well is a genuine two-dimensional problem, just like its commutative counterpart.
Indeed, it can be shown (see in the following) that the Hamiltonian (\ref{schrod-NC-cartes2})
is invariant under passive rotations around the axis of the gravitational field,
which, in our case, was taken to coincide with the $Oy$-axis of the coordinate system. Because of this symmetry, there is no loss of 
generality if we consider, for simplicity, that the initial momentum of the particle has zero component along one of the horizontal axes.
In order to verify this we recall that the effect of a rotation
$\,R(\beta)\,$ of the observer around the $Oy$-axis is given by (in our convention a positive rotation angle $\beta$
corresponds to a counterclockwise rotation of the observer around the $Oy$-axis)
\be
\hat{x}_{1}^{\prime}=
\cos(\beta) \hat{x}_{1} - \sin(\beta) \hat{x}_{3} \,,\quad \hat{x}_{2}^{\prime}=\hat{x}_{2}\,,\quad
\hat{x}_{3}^{\prime}=\sin(\beta) \hat{x}_{1} + \cos(\beta)\hat{x}_{3}\,,
\ee
while the matrix $\theta_{ij}$ transforms as a tensor, that is, 
$\theta_{ij}^{'}=R_{ik}(\beta)\,R_{jl}(\beta)\,\theta_{kl}=R_{ik}(\beta)\,\theta_{kl}\,R_{lj}^{-1}(\beta)$. 
By taking into account the momentum operators transformation rule under a passive rotation around the $Oy$-axis, 
\be \label{momenta-transform}
p_{i}^{\prime}=e^{-i\frac{\beta}{\hbar}L_{y}}\,p_{i}\,e^{i\frac{\beta}{\hbar}L_{y}}\Rightarrow\left\{
\begin{array}{lll}
p_{x}^{\prime}&=&\cos(\beta)p_{x}-\sin(\beta)p_{z}\,,\\
p_{y}^{\prime}&=&p_{y}\,,\\
p_{z}^{\prime}&=&\sin(\beta)p_{x}+\cos(\beta)p_{z}\,,
\end{array}\right.
\ee
one can directly verify that the transformed Hamiltonian $\,H^{\prime}\,$ is such that $\,H^{\prime}=H\,$, as stated.

From the above, we can assume, for simplicity, that the motion takes place in a plane parallel to the $\,xy$-plane, 
without any loss of generality. We will make use of this fact in Sec. \ref{dominios}.

Finally, we remark that we are simply studying the quantum mechanics of a particle with non-commutative spatial 
degrees of freedom and subject to a Newtonian gravitational field. Of course, in principle, one might consider even 
non-commutative corrections to general relativity, but these are expected to be of order $\,\theta_{ij}^2\,$, whereas NCQM 
generally leads to leading corrections of first order in $\,\theta_{ij}\,$. For example, in 
Ref. \cite{Harikumar:2006xf} a non-commutative version of the coupling
of classical gravity to a classical test particle was studied and the leading-order non-commutative correction to the Newtonian potential,
in the linearized approximation, was shown to have the form $\,V_{nc} = V_{Newton}+O(\theta_{ij}^2)\,$. This fact
is specially interesting for us, as it indicates that the non-commutative effects 
coming from the deformation of general relativity and the deformation of the quantum dynamics
that we are considering in the present paper do not mix each other.

\subsection{Domains, self-adjointness, and Hamiltonian spectrum}\label{dominios}

From now on we assume that the motion of the particle takes place in a $z=constant$ plane, so that 
$\,\psi\,$ is a function of the  $\,x\,$ and $\,y\,$ variables solely. The only component of the non-commutativity that matters to our purposes is
$\,\theta_{12}\,$. Hence, in what follows we will simplify the notation, referring to
$\,\theta_{12}\,$ simply as $\,\theta\,$. Thus,  Eq. (\ref{schrod-NC-cartes2}) reduces to
\be
H\psi=-\frac{\hbar ^{2}}{2m}\frac{\partial^{2}\psi}{\partial x^{2}}
-\frac{\hbar ^{2}}{2m}\frac{\partial^{2}\psi}{\partial y^{2}}+mgy\psi - \frac{mgi\theta}{2}\frac{\partial\psi}{\partial x}=E\psi\,\,.
\label{a-hamilt}
\ee 
As it stands, the Schr\"odinger equation, Eq. (\ref{a-hamilt}), is purely formal. 
In order to properly address the energy eigenvalue problem, one must define the domain of 
the Hamiltonian operator $\,H\,$.

Since the formal expression of $\,H\,$ involves differentiation in the \emph{strong} 
sense, that is, in the sense of advanced calculus,
it is not well-defined on the whole Hilbert space $\mathcal{H}$. In such a case we have to restrict the action of $\,H\,$ to a
dense \cite{dense} 
subset, $\,D(H)\subset\mathcal{H}\,$, which we shall call the domain of $\,H\,$ (operators defined on dense domains are called densely defined operators). 
The restriction to dense subsets guarantees the existence of the adjoint \cite{adjointop}  operator, a necessary condition for
one to be able to talk about the self-adjointness of an operator.
We point out that the operators which can be defined on the whole Hilbert space and satisfy  
$\,(T\phi, \psi ) = (\phi, T\psi)\,,\,\,\forall \,\,\phi,\psi\in D(T)\,$ (i.e., that are Hermitian) are necessarily bounded \cite{Reed:1980}. 
On the other hand, an unbounded operator cannot be defined as a self-adjoint operator for all vectors of the Hilbert space. 
It turns out that the Hamiltonian (\ref{a-hamilt}) is an unbounded operator, since $\,D(H)\,$ is a proper subset of $\,\mathcal{H}\,$.


As the configuration space of a particle trapped inside the 
gravitational quantum well is $\,Q=\mathbb{R}\times\mathbb{R}_{+}^{*}\,$, its Hilbert space of states
is $\,\mathcal{H}=L^2(Q)= L^2(\mathbb{R})\otimes L^{2}(\mathbb{R}_{+}^{*})\,$. The domain of the Hamiltonian is 
$\,D(H)=C^{\infty}_{0}(\mathbb{R})\otimes C^{\infty}_{0}(\mathbb{R}_{+}^{*})\,$, 
where $\,C^{\infty}_{0}(\mathbb{R})\,$ denotes the space of functions $\,\phi : \mathbb{R} \to \mathbb{C}\,$ such that $\,\phi\,$ is
infinitely differentiable and have compact support, which means that the set of points where $\,\phi\,$ is not zero is a 
bounded and closed subset of $\,\mathbb{R}\,$ (the space $\,C^{\infty}_{0}(\mathbb{R}_{+}^{*})\,$ is defined in an analogous way).
%
%
Of course, the choice of such a $\,D(H)\,$ was motivated by the fact that
$\,H\,$ is naturally splitted into two independent parts, i.e.,
$\,H=H_{x}\otimes \mathbbm{1}_{D(H_{y})} + \mathbbm{1}_{D(H_{x})}\otimes H_{y}\,,$ where
\be
H_{x}=-\frac{\hbar ^{2}}{2m}\frac{\partial^{2}}{\partial x^{2}}- 
\frac{mgi\theta}{2}\frac{\partial}{\partial x}\,,\qquad 
H_{y}=-\frac{\hbar ^{2}}{2m}\frac{\partial^{2}}{\partial y^{2}}+ mgy\,,
\label{hamilt-split}
\ee
$D(H_{x})=C^{\infty}_{0}(\mathbb{R})\,$ and $\,D(H_{y})=C^{\infty}_{0}(\mathbb{R}_{+}^{*})\,$.

We now investigate the self-adjointness of the Hamiltonian in its domain. By definition, 
an operator is self-adjoint if it is equal to its adjoint. That means that $\,D(T)=D(T^{*})\,$ and 
$\,(T\phi,\psi)=(\phi,T\psi)\,$, for all $\,\phi , \psi \in D(T)\,$. 
Therefore, the self-adjointness of a densely defined operator is equivalent to 
Hermiticity
plus the equality  $\,D(T)=D(T^{*})\,$. Only when an operator is bounded is that Hermiticity implies self-adjointness.
In order to verify the Hermiticity of $\,H\,$ we perform
integration by parts and make use of the reality of $\,V(y)\,$ and the fact that the functions belonging to $\,D(H)\,$ vanish at the boundaries
(they have compact support).
Even though  the Hamiltonian is Hermitian, it turns out that it is not
self-adjoint, since the domain of $\,H^{*}\,$ is larger than the domain of $\,H\,$. Thus $\,H^{*}\,$ is a non-trivial
extension of $\,H\,$, i.e., $\,D(H)\subset D(H^{*})\,$ and $\,H^{*}\phi=H\phi\,$, for all $\,\phi\in D(H)\,$.
In what follows we verify these statements. 

The structure of the Hamiltonian (including the product structure of its domain) means that
the action of $\,H\,$ on elements of $\,D(H)\,$ can be found once we know the actions of 
$\,H_{x}\,$ and $\,H_{y}\,$ into their respective domains. Besides, the adjoint of the Hamiltonian can be expressed as
a linear combination of $\,H_{x}^{*}\,$ and $\,H_{y}^{*}\,$, each one of which can be found separately. 
We consider $\,H_{x}\,$ at first. It is a linear combination of the differential operators
$\,p_{x}:=-i\partial_x\,$ and $\,h_{x}:=-\partial^2_x\,$, both defined on $\,C^{\infty}_{0}(\mathbb{R})\,$.
By definition, if $\,\psi\in D(p_{x}^{*})\,$ then 
$\,(p_{x}\phi , \psi)=(\phi , p_{x}^{*}\psi)\,$, for all $\,\phi\in C^{\infty}_{0}(\mathbb{R})\,$. Therefore,
\be
\int_{\mathbb{R}}dx\,\overline{\partial_x \phi(x)}\,\psi(x) = 
-\int_{\mathbb{R}}dx\,\overline{\phi(x)}\,\left(ip_x^{*}\psi\right)(x),
\ee
for all $\,\phi\in C^{\infty}_{0}(\mathbb{R})\,$. According to the definition of \emph{weak}
derivative \cite{weakder},
it follows that
$\,(p_{x}^{*}\psi)(x)=-i\psi^{'}(x)\,$, where $\,\psi^{'}(x)\,$ is the weak derivative of $\,\psi\in D(p_{x}^{*})\,$ 
with respect to $\,x\,$. 
Besides, since the range of $\,p_{x}^{*}\,$ is a subset of $\,L^{2}(\mathbb{R})\,$,  
it results that
\be
D(p_{x}^{*})=\left\{\psi\in L^2(\mathbb{R}) : \psi^{'}\in L^{2}(\mathbb{R})\right\}.
\label{def-sobolev-1}
\ee
Thus, we may write $\,D(p_{x})=C^{\infty}_{0}(\mathbb{R})\subset D(p_{x}^{*})\,$, which means that
the domain of $\,p_{x}\,$ is a proper subset of the domain of its adjoint.

Regarding $\,h_{x}^{*}\,$, the defining equation is 
$\,\left(h_x \phi, \psi\right)=\left( \phi , h_{x}^{*}\psi\right)\,$, which can be written as
\be
\int_{\mathbb{R}}dx\,\overline{\partial^{2}_x \phi(x)}\,\psi(x) = 
\int_{\mathbb{R}}dx\,\overline{\phi(x)}\,\left(-h_x^{*}\psi\right)(x),
\ee
for all $\,\phi\in C^{\infty}_{0}(\mathbb{R})\,$. 
It results that $\,(h_{x}^*\psi)(x) = -\psi^{''}(x)\,$,
where $\,\psi^{''}(x)\,$ is the second weak derivative of $\,\psi\in D(h_{x}^{*})\,$ with respect
to $\,x\,$. Since
the range of $\,h^{*}_{x}\,$ is a subset of $\,L^{2}(\mathbb{R})\,$, it follows that $\,\psi^{''}\,$ is square integrable. 
Thus, we can write
\be
D\left(h_{x}^{*}\right)=\left\{\psi\in L^2(\mathbb{R}) : \psi^{''}\in L^{2}(\mathbb{R})\right\},
\label{domain2}
\ee
so that
$\,D(h_{x})=C^{\infty}_{0}(\mathbb{R})\subset D(h_{x}^{*})\,$, which means that the domain of $\,h_{x}\,$ is a proper 
subset of the domain of its adjoint. 

According to (\ref{def-sobolev-1}) and (\ref{domain2}), neither $\,p_{x}\,$ nor $\,h_{x}\,$ are self-adjoint in their respective domains,
implying that $\,H_{x}\,$ is not self-adjoint too. Nevertheless, what is important to notice is that 
both $\,p_{x}^{*}\,$ and $\,h_{x}^{*}\,$ are self-adjoint.
Let us take $\,p_{x}^{*}\,$, for example. If $\,\tilde{\psi}(p)=(\mathcal{F}\psi)(p)\,$ denotes the Fourier transform of
$\,\psi\in D\left(h_{x}^{*}\right)\,$, we may write the following familiar results, which can be demonstrated by means of
the usual Fourier transform techniques \cite{Reed:1980,Oliveira:2008},
\be
(\mathcal{F}p_{x}^{*}\psi)(p)=p\tilde{\psi}(p)\,,\qquad
(\mathcal{F}p_{x}^{*}\,\mathcal{F}^{-1}\tilde{\psi})(p)=p\tilde{\psi}(p)\,.
\label{usando-plan}
\ee
We know from Eq. (\ref{def-sobolev-1}) that $\,p_{x}^{*}\psi\,$ is square integrable.
Since the Fourier transform is a unitary 
map from $\,L^{2}(\mathbb{R})\,$ into itself (see \cite{Reed:1980}), 
it follows that $\,p\tilde{\psi}(p)\,$ is square integrable too. Moreover, 
Eq. (\ref{usando-plan}) implies that the operator $\,p_{x}^{*}\,$ is unitarily equivalent
to the operator  $\,\mathcal{M}_{p}\,$ of multiplication by $\,p\,$.
Since the multiplication operator is self-adjoint
in  $\,D(\mathcal{M}_{p})=\lbrace\psi\in L^{2}(\mathbb{R}) : \mathcal{M}_{p}\psi\in L^{2}(\mathbb{R})  \rbrace\,$, then 
$\,p_{x}^{*}\,$ is self-adjoint too. We may write $\,p_{x}^{*}=\mathcal{F}^{-1}\mathcal{M}_{p}\,\mathcal{F}\,$.
A completely anagolous result is valid for $\,h_{x}^{*}\,$, with $\,D(p_{x}^{*})\,$,
$\,\mathcal{M}_{p}\,$ and $\,D(\mathcal{M}_{p})\,$ replaced by $\,D(h_{x}^{*})\,$, $\,\mathcal{M}_{p^2}\,$ and 
$\,D(\mathcal{M}_{p^2})\,$, respectively. It results that by an appropriate choice of its domain, $\,D(H_{x}^{*})\,$, the formal differential operator
\be \label{formal-def}
H_{x}^{*}=\frac{\hbar ^{2}}{2m}h_{x}^{*}+
\frac{mg\theta}{2}p_{x}^{*}
\ee
can be made self-adjoint. Let $\,D_0=D\left(h_{x}^{*}\right)\,\cap\,D\left(p_{x}^{*}\right)\,$. The operator $\,H_{x}^{*}\,$ is 
Hermitian in $\,D_0\,$, but it turns out that this domain is unnecessarily restrictive ($\,D_0\subset D(H_{x}^{*})\,$). In order to characterize $\,D(H_{x}^{*})\,$ we make use of some results from the theory of partial differential equations, notably the Sobolev embedding theorem \cite{Oliveira:2008}.

From the above results we know that if $\,\psi\in D_0\,$ then $\,\psi, \psi^{'}, \psi^{''}\in L^{2}(\mathbb{R})\,$, which means that $\,\psi\,$ belongs to the Sobolev space $\,W^{2,2}=H^{2}\,$, the space of the square integrable functions with square integrable weak derivatives of first and second order. Therefore, according to the Sobolev embedding theorem \cite{Oliveira:2008} $\,\psi\,$ is a continuously differentiable function, that is, $\,\psi\in C^{1}(\mathbb{R})\,$.

Now we make use of the characterization of Sobolev spaces in terms of the absolute 
continuity. A function $\,f : \mathbb{R} \to \mathbb{C}\,$ is absolutely continuous in $\,I=[a,b]\,$ iff there is an integrable function $\,g : I \to \mathbb{C}\,$ such that $\,f(x)=f(a)+\int_{a}^{x}d\xi\,g(\xi)\,\,,\,\,\forall \, x \in [a , b]\,$. If $\,f\,$ is absolutely continuous in every $\,[a,b]\subset\mathbb{R}\,$ then we say that $\,f\,$ is locally absolutely continuous in $\mathbb{R}$ and write $\,\psi \in AC(\mathbb{R})\,$.
This characterization of Sobolev spaces involves the notion of ``absolute continuity on lines'', but in the particular case of functions of one real variable we only need the notion of local absolute continuity (see \cite{hajaz} for details). In order to be allowed to use this result we suppose that $\,\psi^{''}=(\psi^{'})^{'}\,$. Then it follows that if $\,\psi \in D_0\,$ then $\,\psi^{'} \in W^{1,2}\,$, the space of square integrable functions with square integrable first order weak derivative. According to \cite{hajaz} we have $\,W^{1,2} = AC(\mathbb{R})\cap L^{2}(\mathbb{R})\,$, so that $\,\psi^{'}\,$ is locally absolutely continuous.

The above discussion reveals the structure of the domain of self-adjointness of $\,H_{x}^{*}\,$. We notice that it is perfectly possible to drop the much more restrictive requirements of square integrability of $\,\psi^{'}\,$ and $\,\psi^{''}\,$,
since only the range of $\,H_{x}^{*}\,$ must be square integrable, that is,
$\,H^{*}_{x}\psi \in L^{2}(\mathbb{R})\,$. It results that the domain of self-adjointness of $\,H_{x}^{*}\,$ can be written as
\be \label{citardominio}
D(H^{*}_{x})=\left\{\psi\in L^2(\mathbb{R})\cap C^{1}(\mathbb{R}) : \psi^{'}\in AC(\mathbb{R}) \,,\,
-\frac{\hbar^2}{2m}\psi^{''}-\frac{mgi\theta}{2}\psi^{'} \in L^{2}(\mathbb{R})\right \}\,.
\ee
For a good introduction on the importance of Sobolev spaces in quantum physics we refer the reader to \cite{Wan:2006}, where 
the conection between Sobolev spaces and absolute continuity is considered too.

Regarding $\,H_{y}^{*}\,$, the defining equation reads
\be
\,\left(-\frac{\hbar^2}{2m}\partial_y^2\phi + mgy\phi , \psi\right) = 
\left(\phi , H_{y}^{*}\psi\right)\,,
\label{prodL2}
\ee
for all $\,\phi\in C^{\infty}_{0}(\mathbb{R}_{+}^{*})\,$ and for any
$\,\psi\in D(H_{y}^{*})\,$. The Eq. (\ref{prodL2}) leads to 
\be
\int_{0}^{\infty}dy\,\overline{\partial_{y}^{2}\phi(y)} \, \psi(y) = \frac{2m}{\hbar^2}
\int_{0}^{\infty}dy\,\overline{\phi(y)}  \left(mgy-H_{y}^{*}\right)\psi(y)\,.
\ee
Therefore, the second weak derivative of $\,\psi\in D(H_{y}^{*})\,$ with respect to $\,y\,$ and the
action of $\,H_{y}^{*}\,$ on $\,\psi\,$ are respectively given by
\be
\psi^{''}(y)=\frac{2m}{\hbar^2}\left(mgy-H_{y}^{*}\right)\psi(y)\,,\qquad
H_{y}^{*}\psi(y)=-\frac{\hbar^2}{2m}\psi^{''}(y)+mgy\psi(y)\,.
\label{second-weak-vert}
\ee
Since the gravitational potential energy is continuous on $\,\mathbb{R}_{+}^{*}\,$, then it is also locally square integrable,
i.e., it is square integrable on every compact subset of $\,\mathbb{R}^{*}_{+}\,$. We express this fact by writing
$\,V(y)\in L^{2}_{loc}(\mathbb{R}_{+}^{*})\,$. 
Therefore, the product $\,V(y)\psi(y)\,$ is locally square integrable too.
Also, the restriction of a square integrable function to a compact subset of
$\,\mathbb{R}_{+}^{*}\,$ is still square integrable,
so $\,H^{*}_{y}\psi\in L^{2}(\mathbb{R}_{+}^{*})\subset L^{2}_{loc}(\mathbb{R}_{+}^{*})\,$.
Using these facts in Eq. (\ref{second-weak-vert}) we see 
that if $\,\psi\in D(H_{y}^*)\,$, then its second weak derivative $\,\psi^{''}\in L^{2}_{loc}(\mathbb{R}_{+}^{*})\,$. 
Therefore, the domain of $\,H_{y}^{*}\,$ reads
\be \label{dominiodeH-y}
D(H_{y}^{*})=\left\{\psi\in L^{2}(\mathbb{R}_{+}^{*}) : \psi^{''}\in L^{2}_{loc}(\mathbb{R}_{+}^{*}) \,,\, 
-\frac{\hbar^2}{2m}\psi^{''}+mgy\psi\in L^{2}(\mathbb{R}_{+}^{*})\right\}\,,
\ee
from what we see that $\,D(H_{y})=C^{\infty}_{0}(\mathbb{R}_{+}^{*})\subset D(H^{*}_{y})\,$. 
Consequently, the operator $\,H_{y}\,$ is not self-adjoint.

The results presented above show that neither $\,H_{x}\,$ nor $\,H_{y}\,$ are self-adjoint. This fact
leads us to consider the self-adjoint extensions of $\,H\,$. The intuitive idea behind the theory of self-adjoint extensions is that
``the larger the domain of a Hermitian operator the smaller the domain of its adjoint" \cite{Oliveira:2008}. 
In fact,
if $\,S\,$ is a Hermitian extension of a Hermitian operator $\,T\,$, then $\,D(T)\subset D(S)\subset D(S^{*})\subset D(T^{*})\,$.
If the action of $\,S\,$ in its domain
is formally given by the same action as $\,T\,$, then the task of finding self-adjoint extensions of $\,T\,$
reduces to the one of appropriately
choosing the domain of the extension, so that $\,D(S)\,$ turns to be equal to $\,D(S^{*})\,$.
We notice that, according to the above chain of inclusions, every Hermitian 
extension of $\,H\,$ will be a restriction of $\,H^{*}\,$, i.e., the domain of every 
self-adjoint extension of $\,H\,$ will be obtained from the domain of $\,H^{*}\,$ itself.

In order to present the theorem that completely classifies the self-adjoint extensions of any Hermitian operator $\,T\,$ 
acting on a Hilbert space $\,\mathcal{H}\,$, we define
the deficiency subspaces of $\,T\,$, denoted by $\,\mathcal{K}_{+}(T)\,$ and $\,\mathcal{K}_{-}(T)\,$, 
as the vector spaces of the square integrable 
solutions of the equations $\,T\psi=i\lambda\psi\,$ and $\,T\psi=-i\lambda\psi\,$, respectively 
(the positive real constant $\,\lambda\,$ was introduced only for dimensionality considerations). 
The deficiency indices are defined by $\,n_{+}=\textnormal{dim}[\mathcal{K}_{+}(T)]\,$ and $\,n_{-}=\textnormal{dim}[\mathcal{K}_{-}(T)]\,$. 
We say that an operator $\,T\,$ is essentially self-adjoint if and only if its 
closure \cite{closedop} 
is self-adjoint. It turns out that an essentially
self-adjoint operator has only one self-adjoint extension, namely, its closure. 
We also remark that every Hermitian operator is closable. We are now ready to enunciate the von Neumann theorem 
(for details on the above definitions and results and for the complete version 
of the von Neumann theorem, see \cite{Reed:1980,Oliveira:2008}):
%
%
%
\begin{teorema}
Let $\,T\,$ be a 
Hermitian operator on a Hilbert space $\,\mathcal{H}\,$ and let $\,\overline{T}\,$ be its closure. Then: \\
{\textnormal (a)} $\,T\,$ is essentially self-adjoint if and only if $\,n_{+}=n_{-}=0\,$; \\
{\textnormal (b)} $\,T\,$ has self-adjoint extensions if and only if $\,n_{+}\,=\,n_{-}\,$; \\
{\textnormal (c)} There is a one-to-one correspondence between the unitary maps
$\,U : \mathcal{K}_{+}(T) \to \mathcal{K}_{-}(T)\,$ and the self-adjoint extensions of $\,T\,$ (which we shall denote by $\,T_{U}\,$);\\
{\textnormal (d)} The domain of $\,T_{U}\,$ is
$\,D(T_{U})=\left\lbrace\psi+\psi_{+}+U\psi_{+}\in D(T^{*}) : \psi\in D(\overline{T})\,,\,\psi_{+}
\in D(U)\right\rbrace$. 
\label{teo}
\end{teorema}

A few remarks on the consequences of the above theorem are worthy. First of all,
we realize that a Hermitian operator may have several self-adjoint 
extensions, as well as it may have none. 
In the former case, the analysis of the physical conditions governing the behavior of the system at the boundaries 
may help one to choose the appropriate self-adjoint extension. We also note that in case $\,n_{+}=n_{-}=n\geq 1\,$, 
any self-adjoint extension of $\,T\,$ corresponds to a unitary $\,n\times n\,$ matrix. That means that the
family of self-adjoint extensions is parametrized by 
$\,n^{2}\,$ real parameters, corresponding to the $\,n^{2}\,$ independent 
parameters of the Lie group $\,U(n)\,$. Finally, we notice that the self-adjoint extensions of a Hermitian operator $\,T\,$
are restrictions of $\,T^{*}\,$ to appropriate subsets of $\,D(T^{*})\,$. Once we know the
formal expression of $\,T^{*}\,$, the task of finding the self-adjoint extensions of $\,T\,$
reduces to that of finding the domains of self-adjointness of the formal expression of $\,T^{*}\,$.

We now apply the above mathematical results to the task of obtaining the self-adjoint extensions of $\,H_{x}\,$, $\,H_{y}\,$ 
and $\,H\,$.
Regarding $\,H_{x}\,$, the relevant differential equations, as well as their solutions, are given respectively by
\be
-\frac{\hbar ^{2}}{2m}\psi_{\pm}^{''}(x)-
\frac{mgi\theta}{2}\psi_{\pm}^{'}(x)=\pm i\lambda\psi_{\pm}(x)\,,\qquad\psi_{\pm}(x)=e^{\omega_{\pm}x}\,.
\ee
The characteristic equation for each one of the complex constants $\,\omega_{\pm}\,$ is a quadratic equation, which can be readily solved.
Hence, all the solutions $\,\psi_{\pm}(x)\,$ are
divergent in one or the other of the end points ($\,\pm \infty\,$)
of the domain of integration.
Consequently, all these functions fail to be square 
integrable on $\,\mathbb{R}\,$. 
Therefore, we have $\,n_{+}=n_{-}=0\,$, which means that
$\,H_{x}\,$ is essentially self-adjoint. Its closure, $\,\overline{H}_{x}\,$, is its unique self-adjoint extension, which 
implies that (it should be noted that if an operator $\,T\,$ and its adjoint $\,T^{*}\,$ are Hermitian operators, then $\,T^{*}\,$ is self-adjoint \cite{Oliveira:2008})
 $\,\overline{H}_{x}=H^{*}_{x}\,$, so we can write
\be
\overline{H}_{x}=\frac{\hbar ^{2}}{2m}h_{x}^{*}+ 
\frac{mg\theta}{2}p_{x}^{*}\,,\qquad D(\overline{H}_{x})=D(H^{*}_{x}),
\qquad\mbox{as given by Eq. (\ref{citardominio})}\,.
\ee

Regarding $\,H_{y}\,$, each one of the equations $\,H_{y}\psi=\pm i\lambda\psi\,$ can be put into the form of an 
Airy equation in the complex plane \cite{Olver:1974}, i.e.,
\be
-\frac{\hbar^{2}}{2m}\psi^{''}_{\pm}(y)+mgy\,\psi_{\pm}(y)=\pm i\lambda\psi_{\pm}(y)\quad\Longrightarrow\quad
\psi^{''}_{\pm}(z_{\pm})-\left(\frac{2m^{2}g}{\hbar^{2}}\right)z_{\pm}\,\psi_{\pm}(z_{\pm})=0\,,
\label{airy-sim}
\ee
where $\,z_{\pm}=y\mp i\lambda/(mg)\,$. The complex Airy equation has two linearly independent complex solutions, the 
Airy functions, $\,Ai(z)\,$ and $\,Bi(z)\,$, so that the differential equation for $\,\psi_{+}\,$ has two linearly independent solutions,  
$\,Ai(z_{+})\,$ and $\,Bi(z_{+})\,$, and analogously for $\,\psi_{-}\,$.
In order to investigate the square-integrability of the wave functions $\,\psi_{\pm}^{a}(y):=Ai(z_{\pm})\,$ and 
$\,\psi_{\pm}^{b}(y):=Bi(z_{\pm})\,$
on $\,\mathbb{R}_{+}^{*}\,$, we look at their behavior 
near the boundaries of the domain of integration. As the potential is well behaved at the origin, we only have to care about the
asymptotic behavior of $\,Ai(z_{\pm})\,$ and $\,Bi(z_{\pm})\,$ for 
$\,|z_{\pm}|\rightarrow\infty\,$ (indeed, for $\,y=\textnormal{Re}(z_{\pm})\rightarrow\infty\,$).
Since $\,\textnormal{Re}(z_{\pm})>0\,$ and $\,\textnormal{Im}(z_{\pm})\,$ is a non-zero complex constant, the complex variables $\,z_{\pm}\,$
belong to the sector of the complex plane defined by $\,|\textnormal{Arg} (z_{\pm})| \le\pi -\delta\,$, for some $\,\delta >0\,$ 
(the variables $\,z_{\pm}\,$ never
reach the negative $Ox$-axis). In this case, the asymptotic expressions read \cite{Abramowitz:1972}
\begin{eqnarray}
Ai(z_{\pm})\,\sim \, \frac{1}{2\sqrt{\pi}}\,\,z_{\pm}^{-1/4}\,e^{-\frac{2}{3}z_{\pm}^{3/2}}\,,\qquad
Bi(z_{\pm})\,\sim \, \frac{1}{\sqrt{\pi}}\,\,z_{\pm}^{-1/4}\,e^{\frac{2}{3}z_{\pm}^{3/2}}\,.
\end{eqnarray}
One can see that, in both cases ($\pm i \lambda\,$), only the solutions $\,\psi^{a}_{\pm}(y)=Ai(z_{\pm})\,$ are square integrable in the domain of 
integration $\,\mathbb{R}_{+}^{*}=(0 , \infty)\,$,
which means that the deficiency indices of the operator $\,H_{y}\,$ are both equal to 1. 
Thus, since $\,n_{+}=n_{-}=1\,$, it follows that the family of self-adjoint extensions of $\,H_{y}\,$ is 
parametrized by just one real number. 

Let $\,S\,$ be a self-adjoint extension of $\,H_{y}\,$, with domain $\,D(S)\subset D(H_{y}^{*})\,$. We know from Theorem \ref{teo} 
(see also the remarks that followed the theorem) that any self-adjoint extension of $\,H_{y}\,$ is a restriction of $\,H_{y}^{*}\,$, i.e., 
$\,S\psi=H_{y}^{*}\psi\,\,,\,\,\forall \psi \in D(S)\,$. 
The domain $\,D(S)\,$ is characterized as a subset of $\,D(H_{y}^{*})\,$
whose elements fulfill a particular set of boundary conditions. In order to find these boundary conditions 
we make use of the Hermiticity of $\,S\,$ and the fact that this operator is a restriction of $\,H_{y}^{*}\,$. As a result, we have
\be \label{condicao-hermiticidade-adjunto}
(\,H_{y}^{*}\phi , \psi)=(\phi,H_{y}^{*}\psi)\,,
\ee
for all $\, \phi , \psi \in D(S)\,$, what leads to (see Eq. (\ref{second-weak-vert}))
\be
\int_{0}^{\infty}dy\,\overline{\psi^{''}(y)} \, \phi(y)=
\int_{0}^{\infty}dy\,\overline{\psi(y)}\,\phi^{''}(y) \,,\qquad\forall\,\,\phi , \psi \in D(S)\,.
\label{calcula-extension}
\ee
Notice that the potential terms were readily cancelled out. Then, 
performing integration by parts on both sides of Eq. (\ref{calcula-extension}) we find
\begin{equation}
\overline{\psi^{'}(\infty)}\,\phi(\infty)-\overline{\psi(\infty)}\,\phi^{'}(\infty)=
\overline{\psi^{'}(0)}\,\phi(0)-\overline{\psi(0)}\,\phi^{'}(0)\,.
\end{equation}
We now make use of the fact that the functions belonging to $\,D(H^{*}_{y})\,$, as defined in Eq. (\ref{dominiodeH-y}),
vanish at infinity (see \cite{Voronov:2008} for the demonstration of this result). It results that
\be
\frac{\overline{\psi^{'}(0)}}{\overline{\psi(0)}}=\frac{\phi^{'}(0)}{\phi(0)}\,.
\label{bound-self-ad}
\ee 
The Eq. (\ref{bound-self-ad}) is fulfilled if we impose the boundary condition 
$\,\psi(0)=\alpha\,\psi^{'}(0)\,$, $\,\alpha\in\mathbb{R}\,$,  for all functions belonging to  
$\,D(S)\,$. This fact leads us to modify the notation used in the above considerations and write $\,H_{y,\alpha}\,$ 
in place of $\,S\,$. Therefore, 
\be \label{dominio-das-extensoes-AA}
D(H_{y,\alpha})=\left\lbrace \psi\in D(H_{y}^{*}) : \psi(0)=\alpha\,\psi^{'}(0)    \right\rbrace.
\ee
The range of $\,\alpha\,$ can be extended so as to comprehend the case $\,\alpha=\infty\,$, which corresponds to $\,\psi^{'}(0)=0\,$. 

The self-adjoint boundary conditions in Eq. (\ref{dominio-das-extensoes-AA}) can be recognized as the canonical self-adjoint boundary conditions of the Sturm-Liouville problem for the operator $\,H_{y}\,$ (see, e.g., \cite{coddington}).


Let us finally calculate the energies of the particle in the gravitational quantum well.
As usual, the energies are given by the eigenvalues of the Hamiltonian operator of the system. But according to 
Eq. (\ref{dominio-das-extensoes-AA}),
there exists a whole family of mathematically allowed self-adjoint Hamiltonians which could 
be associated to a quantum particle moving under the action of a uniform gravitational potential.

The infinite forces  experienced by the particle at the perfectly reflecting mirror, situated at the bottom of the well, 
lead us to impose the condition $\,\psi(0)=0\,$. This boundary condition corresponds to the choice  $\,\alpha=0\,$ in Eq. (\ref{dominio-das-extensoes-AA}).
Therefore, the natural boundary condition suggested by the analysis of the physics of the problem has selected one of the possible
self-adjoint extensions of the Hamiltonian $\,H_{y}\,$, and therefore of the full Hamiltonian $\,H\,$. The other self-adjoint 
extensions obtained  in Sec. (\ref{dominios}) correspond to situations in which partial reflection occurs at the mirror. 
The boundary conditions corresponding to those extensions may be relevant to further investigations on the phenomenology of the 
gravitational quantum well.

We can now proceed to solve the energy eigenvalue problem. 
Since the motion is
free along $\,Ox\,$, we use the {\it ansatz} $\,\psi (x,y)=e^{ikx}\phi (y)\,$ into (\ref{a-hamilt}) to get
\be
\frac{d^{2}\phi}{dy^{2}} +\frac{2m^{2}g}{\hbar ^{2}}
\left (\frac{\mathcal{E}_{\theta}}{mg}-y\right)
\phi(y)=0\,,
\label{eq-dif1}
\ee
where 
\be
\mathcal{E}_{\theta} = E-\frac{\hbar ^{2}k^2}{2m}-\frac{mgk\theta}{2}\,.
\label{E_theta}
\ee

Now, by setting
\be \label{definicoes}
\xi =\frac{y-b_{\theta}}{a}\,, \qquad a=\left (\frac{\hbar ^{2}}{2m^{2}g}\right ) ^{1/3}\,, \qquad 
b_{\theta}=\frac{\mathcal{E}_{\theta}}{mg}\,, 
\ee 
we can put (\ref{eq-dif1}) in the form of an Airy 
equation
\be
\frac{d^{2}\phi(\xi)}{d\xi ^{2}} -\xi\phi(\xi)=0\,, 
\label{eq-dif2}
\ee 
whose general solution is
\be
\phi(y)=A\,Ai\left (\frac{y-b_{\theta}}{a}\right )
+B\,Bi\left( \frac{y-b_{\theta}}{a}\right )\,.
\label{soluc-y}
\ee
We now simply impose the adequate boundary conditions for $\,\phi(y)\,$,
\begin{equation}
\phi(0)=0\,,\qquad
\lim_{y\rightarrow\infty}\phi(y) = 0\,,
\end{equation}
what leads to 
\be
\phi(y)=A\,Ai\left(\frac{y-b_{\theta}}{a}\right).
\label{soluc-y2}
\ee

The application of the boundary condition of vanishing 
wave-function at the mirror leads to
$\,Ai \left (-\frac{b_{\theta}}{a}\right)=0\,$,
so that $\,-b_{\theta}/a\,$ are the roots of the Airy function $\,Ai\,$, i.e.,
\be
b_{\theta,n}=-a\,\alpha _n\,,
\label{zeros-airy}
\ee
where $\,\alpha _n\,$ denotes the $\,n\,$-th zero of $\,Ai\,$. The result (\ref{zeros-airy})
combined with the definition of $\,b_{\theta}\,$ (see Eq.s (\ref{E_theta}) and 
(\ref{definicoes})) gives the spectrum of the Hamiltonian
\be
E_{k,n,\theta} = \frac{\hbar^2 k^2}{2m}\,+\,\left
(\frac{mg^2 \hbar ^2}{2}\right )^{1/3}.\,(-\alpha _n)\,+\,
\frac{mgk\theta}{2}\,.
\label{espectro-energ}
\ee
For $\,\theta =0\,$, the result found in ordinary quantum mechanics is recovered. We note that the effect of non-commutativity also 
disappears for $\,k=0\,$, what simply corresponds to a one-dimensional movement, along 
the direction of the gravitational field. 

We remark that the energy spectrum of the non-commutative gravitational quantum well is invariant under parity. 
Indeed, recall that in two dimensions the matrix element of the parity operator 
is given by $\,P_{ij}=-\epsilon_{ij}\,$, where $\,\epsilon_{ij}\,$ is the Levi-Civita symbol. Thus the non-commutativity matrix
transforms as $\theta_{ij}^{\prime}=-\theta_{ij}\,$, that is, $\theta^{\prime}=\theta^{\prime}_{12} = -\theta\,$. On the other hand, 
since the momentum operator $\,\hat{p}_{x}\,$ anti-commutes with 
$\,P\,$, it follows that the parity transformation implies that
$\,k\,$ goes into $\,-k\,$, thus showing that the spectrum in Eq. (\ref{espectro-energ}) is invariant under $\,P\,$.
It is important to notice that if we had adopted the three dimensional scenario for the problem of the quantum well, we would have 
found the same result, since the parity transformation in three dimensions
changes the sign of $\,g\,$, while keeping $\,\theta_{ij}\,$ unchanged. In any case the expression of the energy is unaffected by $\,P\,$,
a result which obviously follows from the fact that $\,H\,$ commutes with $\,P\,$.

We finish this section by analysing what is, now in the non-commutative case,
the analogue of the classical turning point. Just like in the ordinary case, it corresponds to the
vertical position from where the wave functions start to fall off exponentially.
From Eq. (\ref{definicoes}) we see that $\,\xi =0\,$ implies
\be \label{define-b}
y= b_{\theta} =\frac{E-\frac{\hbar^2 k^2}{2m}-\frac{mgk\theta}{2}}{mg}=b-\frac{k\theta}{2}\,,
\ee
where $\,b\,$ denotes the classical turning point.
Thus we see that for $\,\theta\neq 0\,$ the turning point position acquires a $\theta$-dependent correction,
a result that is compatible with the corresponding non-commutative deformation of the classical equations of motion.

\section{The GRANIT experiment: an upper-bound on $\,\theta$}\label{granit-section}

Having found the bound states of a particle in a non-commutative 
gravitational quantum well, we can now constrain the value of the 
non-commutative parameter, by resorting to data of the 
experiment performed by Nesvizhevsky et al. - the GRANIT 
experiment \cite{Nesvizhevsky:2002,Nesvizhevsky:2005ss}. In this 
experiment the lowest quantum bound states of neutrons on free-fall 
in the Earth's gravitational field were observed and their energy determined. 
Due to the weaker strength of gravity, as compared to nuclear and electromagnetic interactions, it is very difficult to perform experiments in which gravity plays a role
in the manifestation of the quantum nature of matter.  Thus the GRANIT experiment, performed under challenging, extremely sensitive experimental designed conditions, is another striking observation of the wave behavior of matter in a gravitational field (an earlier experiment in 
which the quantum, wave nature of matter was manifested, due to its interaction 
with the Earth's gravitational field, is that of the observation of gravitationally 
induced quantum interference of neutrons \cite{Werner:1975wf}. The data of this experiment could also be used to constrain the value of $\theta\,$). This was achieved by means of a
 gravitational quantum well, formed by the Earth's gravitational field and a
 horizontal reflecting mirror (considered as perfectly reflecting). 
Due to their charge neutrality, long lifetime, and low mass, neutrons 
are suitable to perform this kind of experiment, in which the effect 
of the interactions other than gravity must be negligible. A horizontal 
beam of ultracold neutrons was thus allowed to fall freely, flying above the 
reflecting mirror at the bottom. As no forces act on 
neutrons horizontally and just gravity acts vertically, we have a
 gravitational potential well along the latter direction. For details on the
 experimental setup, see 
\cite{Nesvizhevsky:2002,Nesvizhevsky:2005ss}. 

An upper-bound on $\,\theta\,$ can be established by imposing 
that the $\,\theta\,$-dependent corrections to the energy implied by Eq. (\ref{espectro-energ}) 
be smaller or of the order of the maximum differences of the energy 
levels provided by the GRANIT experiment, i.e., according to 
its error bars. Hence we require that 
\be 
\frac{mgk\theta}{2}\lesssim\Delta E^{exp}\,, \ee so that \be
\theta\lesssim\frac{2\Delta E^{exp}}{mgk}\,. 
\label{bound} 
\ee

For the first two energy levels, the experiment gives 
$\,\Delta E_1^{exp} = 6.55\times 10^{-32}\,J\quad (n=1)\,$ and $\,\Delta
E_2^{exp}=8.68\times 10^{-32}\,J\quad (n=2)\,$. For neutrons we 
have $\,m\simeq 1.675\times 10^{-27}\,kg\,$, and in the experiment the
 neutrons 
had a mean horizontal speed $\,<v_x>\,\simeq 6.5\,m/s\,$
 \cite{Nesvizhevsky:2005ss}, so that
$\,k=<p_x>/\hbar =m<v_x>/\hbar\simeq 1.03\times 10^{8}\,m^{-1}\,$.
Then, by considering $\,g\simeq 9.81\,m/s^{2}\,$ we arrive at the following
upper-bounds for $\,\theta\,$ 
\bear 
\theta & \lesssim &
0.771\times 10^{-13}\,m^{2}\qquad (n=1)\,,\label{bound1}\\
\theta & \lesssim & 1.021\times 10^{-13}\,m^{2}\qquad (n=2)
\,.
\label{bound2} 
\eear

We point out that the literature on the non-commutative gravitational well that resorts
to data of the GRANIT experiment do not arrive at un upper-bound on $\,\theta\,$. 
As they considered non-commutativity in both configuration and momentum
 spaces, they have found an upper-bound on the
parameter of momentum non-commutativity instead
 \cite{Bertolami:2005jw}-\cite{Buisseret:2007qd}.  
Since we consider spatial non-commutativity only, 
we can establish an upper-bound on the parameter of spatial 
non-commutativity. 
We remark that other upper-bounds on $\,\theta\,$ were established by 
experimental data related to the Lorentz invariance 
\cite{Carroll:2001ws} and to the Lamb shift 
\cite{Chaichian:2000si}, for instance. 
 
One of the most important prospects about the GRANIT experiment is 
the improvement of the energy resolution of the neutrons energy 
levels. In principle, with the energy resolution, $\,\Delta E\,$, 
limited only by the uncertainty principle ($\Delta E\,\Delta t\,\thicksim \hbar\,$), 
one could achieve a value as low as $\,10^{-18}eV\,$, if $\,\Delta t\,$ approaches the neutron 
lifetime. Improvement in the energy resolution could help in 
testing the proportionality between inertial and gravitational 
masses for neutrons \cite{Nesvizhevsky:2005ss}. 
The Equivalence Principle in the context of the gravitational quantum well was studied in 
\cite{Bertolami:2002xh}, but in a manner completely different from that we consider 
in the following section. In fact, in the following, we do not need even 
to consider that a well is set up, the Equivalence Principle being studied by means of 
the time of flight of a quantum particle.

\section{Quantum clock and time of flight measurement}\label{time_section}

By resorting to the basic notion that ``a clock is a dynamical system 
which passes through a succession of states at constant time intervals'' 
\cite{Peres:1980}, that is, that ``... the measurement of 
time actually is the observation of some dynamical variable, the law 
of motion of which is known ...'' \cite{Peres:1980}, Peres 
has modeled a simple quantum clock by means of a quantum rotor 
(also known as the Larmor clock). 
When coupled to another system, this can measure the duration of
 physical processes 
as well as keep a permanent record of it, such as the time of flight of
 a particle, 
or even to control the duration of a physical process \cite{Peres:1980} (see
 also \cite{Peres:1993}). 

In order to address the essential features involved in modeling a
 quantum clock, and mainly to prepare the discussion in Sec. \ref{sec_tempo}, 
in the following we briefly review Peres construction of a quantum
 clock and its particular use in a time of flight experiment involving 
the motion of a free quantum particle, as considered in \cite{Peres:1980}.

Peres basic ideia in modeling a quantum clock is to
 consider a quantum rotor, since this can be regarded as describing the
 motion
 of a pointer on a clock dial. Thus, he considered 
a clock with an odd number of ``pointer states'', $\,N=2j+1\,$,
 represented by 
\begin{equation}
u_n(q)=\frac{1}{\sqrt{2\pi}}e^{inq}\,,\qquad n=-j,...,j\,,
\label{base1}
\end{equation}
where $\,q\,$ is the clock's degree of freedom and $\,0\leq q < 2\pi\,$. 

Another suitable orthogonal basis of states for the clock is 
\begin{eqnarray}
v_\kappa &=&\frac{1}{\sqrt{N}}\sum_{n=-j}^{j}\,e^{-i2\pi\kappa n/N}\,u_n (q)\,,\nonumber\\
&=& \frac{1}{\sqrt{2\pi N}}\frac{\sin\frac{N}{2}(q-\frac{2\pi\kappa}{N})}
{\sin\frac{1}{2}(q-\frac{2\pi\kappa}{N})}\,,
\label{base2}
\end{eqnarray}
where $\,\kappa =0,...,N-1\,$.
It follows that for large values of $\,N\,$ the state $\,v_\kappa (q)\,$ 
is sharply picked at $\,2\pi\kappa/N\,$. 
We can refer to the states $\,v_\kappa$ as the clock states.

By defining the projection operators $\,P_\kappa v_n ={\delta}_{\kappa n} v_n\,$,
a ``clock time'' operator can be defined as
\begin{equation}
T_c  =\tau{\sum _{\kappa}}\kappa P_\kappa \,, 
\end{equation}
where $\,\tau\,$ is the time resolution of the clock. 
Its eigenstates are $\,v_\kappa\,$, with $\,t_\kappa=\kappa\tau\,$  
as the corresponding eigenvalues.

The clock's Hamiltonian is 
\begin{equation}
H_c = \omega J\,,\qquad\omega =2\pi/(N\tau)\,,\qquad J=-i\hbar\partial /\partial q\,.
\label{hamilton-relogio}
\end{equation}
The eigenstates of $\,H_c\,$ are the vectors $\,u_n(q)\,$, defined in Eq. (\ref{base1}).
The eigenvalue corresponding to $\,u_{n}(q)\,$ is $\,n\hbar\omega\,$. 
From (\ref{base1}) and (\ref{hamilton-relogio}) it follows that 
\begin{equation}
e^{-iH_c\tau /\hbar}v_\kappa (q) =v_{\kappa +1(mod\,N)}(q)\,.
\end{equation}
Assuming that $\,v_0\,$ is the initial state of the clock, the above result implies
 that the 
clock will pass successively through the states $\,v_0,v_1,v_2,...\,$, at
 time 
intervals given by $\,\tau\,$.

Before we apply the Peres quantum clock model to our case of interest, firstly it is very 
intructive to recall its application in determining  the time of fight of a free particle traveling between two assigned points
\cite{Peres:1980}. One has to demand that the clock be activated when the
 particle pass by the first 
point and then be stopped when it passes by the second one. Let us consider that the
 particle moves along 
the direction $\,Ox\,$, so that $\,x_1\,$ and $\,x_2\,$ correspond to the two 
positions of interest. 
In the Schr\"{o}dinger representation the Hamiltonian for the composed, 
particle-clock system reads
\begin{equation}
H= -\frac{\hbar ^{2}}{2m}\frac{\partial^{2}}{\partial x^{2}}-i\hbar\omega\Theta
 (x-x_1)\Theta (x_2-x)
\frac{\partial}{\partial q}\,,
\label{hamiltoniana-peres}
\end{equation}
so that the clock runs only if the particle lies within the interval $\,[x_{1} , x_{2}]\,$.

Following Peres, we set the initial state of the clock as $\,v_0 ={\sum _{n}}
u_n/\sqrt {N}\,$. As $\,H\,$ and
 $\,H_c =\omega J\,$ 
commute, it is simpler to solve the equation of motion for the clock in
 an eingenstate 
of $\,J\,$, i.e., $\,u_n \,$, and then sum over these partial solutions to get the evolved wave function corresponding to the 
initial state, $\,v_0\,$. Therefore, the initial state of the 
(still not coupled) particle-clock 
system can be written as $\,\Psi_0 (x,q)e^{-iEt/\hbar}\,$, with 
$\,\Psi_0 (x,q)\,$ 
given in the factorized form (see \cite{Peres:1980,Leavens:1993})
\begin{equation}
\Psi_0 (x,q)=A_k\,e^{ikx}v_0 (q)=A_k\,
 e^{ikx}\frac{1}{\sqrt{N}}\sum_{n=-j}^j u_n (q)\,,
\label{ondainicial}
\end{equation}
where $\,E\,$ is the energy of the particle, $\,k=\sqrt{2mE}/\hbar\,$ and $\,A_k\,$ is a normalization constant.

On the other hand, the final state of the system cannot be 
factorized as 
$\,\Psi_0 (x,q)\,$, since the motion of the particle in the region $\,x_1 <
 x < x_2\,$ will activate the clock, making the particle 
and clock coordinates to mix. Thus we write 
\begin{equation}
\Psi (x,q)= \frac{1}{\sqrt{N}}\sum_{n=-j}^j \psi^{n}_k
 (x)u_n (q)\,.
\label{ondafinal}
\end{equation}

The substitution of (\ref{ondafinal}) into the Schr\"{o}dinger
 equation
 leads to
\begin{equation}
-\frac{\hbar ^{2}}{2m}\frac{d^{2}\psi^{n}_k}{dx^{2}}+ \biggl[
 n\hbar\omega\Theta (x-x_1)\Theta (x_2-x)
-E\biggr] \psi^{n}_k (x) = 0\,.
\end{equation}

Thus, we see that $\,\psi^{n}_k\,$ satisfies the 
Schr\"{o}dinger equation for a particle under the action of a rectangular 
barrier of height 
$\,V_n = n\hbar\omega\,$ and length $\,L=x_2 -x_1\,$. 
Outside the barrier we have (we are now basically 
following Leavens presentation of the original Peres quantum clock 
model \cite{Leavens:1993})  
\begin{equation}
\psi^{n}_k (x) = \left\{
\begin{array}{cccc}
& e^{ikx}+R_{n}(k)\,e^{-ikx} &,& \qquad  x<x_1\,,\\
& T_{n}(k)\,e^{ikx}&,& \qquad x>x_2\,,
\end{array}\right.
\end{equation}
where $\,T_{n}(k)\,$ and $\,R_{n}(k)\,$ are the transmission and reflexion amplitudes, respectively, given by
\begin{equation}
T_{n}(k) = \frac{e^{-i k L}}{\cos{(k_{n}L)}-\frac{i}{2}\left(\frac{k}{k_{n}}+\frac{k_{n}}{k}\right)\sin{(k_{n}L)}}
\label{amplit_trans}
\end{equation}
and 
\begin{equation}
R_{n}(k) =
\frac{\frac{i}{2}\left(\frac{k_{n}}{k}-\frac{k}{k_{n}}\right)e^{2ikx_{1}}\sin{(k_{n}L)}}{\cos{(k_{n}L)}-\frac{i}{2}\left(\frac{k}{k_{n}}+\frac{k_{n}}{k}
\right)\sin{(k_{n}L)}}\,,
\label{amplit_refl}
\end{equation}
and where $\,k_{n}=\sqrt{2m(E-n\hbar\omega)}/\hbar\,$.

Of course that the particle will be perturbed when coupled 
to the physical clock. In order that one has a small disturbance 
on the particle evolution, one must consider that such 
coupling is sufficiently small. We therefore set $\,V_n \ll E\,$, 
from which we have 
\be
k_n \simeq k-n\omega (2E/m)^{-1/2}=k-n\omega /(\hbar k/m)\,.
\ee 
Furthermore, in the limit of small coupling we also have $\,T_{n}(k) \simeq\exp{(i(k_{n}-k)L)} \,,$
so that $\,|T_{n}(k)|\simeq 1\,$.
Under the above conditions we can write the phase shift due to the clock as
\begin{equation}
(k_n - k)L \simeq -n\omega L(2E/m)^{-1/2}=-\frac{n\omega L}{\hbar
 k/m}\,.
\end{equation}

Let us call $\,T=L/(\hbar k/m)\,$.
Now we can express the final wave function for the 
particle-clock system as  
\begin{equation}
\Psi (x,q)\simeq A_k
 e^{ikx}\frac{1}{\sqrt{N}}\sum_{n=-j}^j e^{-in\omega
 L/(\hbar k/m)}
u_n (q)\,,
\label{tempo}
\end{equation}
that is, 
\begin{equation}
\Psi (x,q)\simeq A_k
 e^{ikx}\frac{1}{\sqrt{N}}\sum_{n=-j}^j e^{-in\omega T}
u_n (q)=A_k e^{ikx}v_0(q -\omega T)\,.
\label{tempo2}
\end{equation}

Since $\,e^{-i\hat{H}_c T/\hbar}u_n (q)= u_n (q -\omega T)\,$,
 it follows that 
the pointer, initially directed to $\,q = 0\,$, will be found
 directed to 
$\,q\simeq\omega T\,$ after the particle leaves the region where 
the clock runs. 
By noting that $\,v=\sqrt{2E/m}=\hbar k/m\,$ is the velocity of a free quantum particle with energy $\,E\,$, we see that 
the Peres clock records the time of 
flight of a particle which travels between two assigned points. The key fact 
is that the time of flight is encoded in the phase shift due to the clock barrier. 

We remark that the quantum clock does not measure the absolute 
instants of time in which the particle passes through the positions 
$\,x_1\,$ and $\,x_2\,$, but only the time difference to travel 
between them. This fact avoids the collapse of the particle wave 
function \cite{Peres:1980}. This is the advantage of 
using a quantum clock to study delocalized states, as energy eigenstates 
(see Sec. \ref{sec_tempo}).

\section{Time of flight in a uniform gravitational field in NCQM} \label{sec_tempo}

\indent

In \cite{Davies:2006} Davies applied the Peres model of a quantum clock to 
determine the time of flight of a quantum particle in its round trip in a uniform 
gravitational field, when it is vertically projected up. Davies' interest 
in studying this sort of ``quantum Galileo experiment'' was motivated by asking 
if there would be a violation of the (Weak) Equivalence Principle 
when one considers quantum particles, as they may tunnel into the classically 
forbidden region of the gravitational potential. 
Therefore, a mass-dependent delay in the time of flight might result, as 
compared to the classical case. 
As Davies remarks, the answer to this question is not {\it a priori} 
obvious in the case one is handling with energy eigenstates, 
which are delocalized states, with no corresponding classical 
counterparts on localized bodies (for a study of the motion of wave packets and 
the ``quantum Galileo experiment'', see \cite{Viola:1996de}). 
Davies finds, by assuming the equality of the inertial ($m_i$) and 
gravitational ($m_g$) masses, that when the measurement is made far 
from the classical turning point, the time of flight is equal to the 
classical result. Therefore, in this sense, the (Weak) Equivalence Principle 
holds in quantum mechanics, even when one considers highly non-classical 
states. 

Inspired by Davies, we next apply the quantum clock model of Peres 
to determine the time of flight of a particle in a uniform gravitational
field in the context of NCQM, in order to address the 
``quantum Galileo experiment'' and the question of the validity 
of the Equivalence Principle in NCQM. We will assume that $m_i=m_g\equiv m\,$.

We now make use of the simple clock model of Sec. \ref{time_section} to measure the
interval of time during which the particle lies within the semi-infinite
region of space defined by $\,y_{0}\leq y <\infty\,$, 
where $\,y_{0}\,$ is some fixed height, at where we can consider that 
the clock is set up. We consider that the particle is obliquely projected up 
from some height below $\,y_{0}\,$.

Let $\,P_{y_{0}}\,$ be the projection operator defined by 
\be
P_{y_{0}}\psi(x , y)=\Theta(y-y_{0})\psi(x , y)\,.
\ee
The vertical particle degree of freedom couples to that of the
 quantum clock through the interaction 
Hamiltonian 
\be
H_{I}=P_{y_{0}}\otimes H_{c}=\Theta(y-y_{0})\omega J\,.
\label{interaction-clock}
\ee
The action of the total Hamiltonian on wave 
functions reads
\be
H\Psi = -\frac{\hbar ^{2}}{2m}\frac{\partial^{2}\Psi }{\partial x^{2}}
-\frac{\hbar ^{2}}{2m}\frac{\partial^{2}\Psi }{\partial y^{2}}
+ mgy\Psi\ +
 \Theta(y-y_{0})  \left(\omega J\Psi \right).
\ee

We now consider the non-commutative analogue of the above problem. We know 
that the action of the potential energy on non-commutative wave functions can be written in terms of the  Moyal product,
$\,V\star\psi\,$. In order to calculate this function we note that there are two different contributions to the
total interaction. For the gravitational potential we can directly
apply Eq. (\ref{Moyal_product_definition}), thus obtaining the following exact result,
\be
mgy\star\Psi (x,y)=mg\left (y\Psi
(x,y)-\frac{i\theta}{2}\frac{\partial\Psi}{\partial x}\right).
\label{moyal-axact-gravit}
\ee

Denoting the energy of the particle-clock system by $\,E\,$, its eigenstate is
given by $\,\Psi(x,y,q)e^{-iEt/\hbar}\,$. From Eq. (\ref{eq-schrodinger-NC-indep-t}) and Eq. (\ref{moyal-axact-gravit}) we thus have 
\be \label{Schroedinger_equation_2}
-\frac{\hbar ^{2}}{2m}\frac{\partial^{2}\Psi }{\partial x^{2}}
-\frac{\hbar ^{2}}{2m}\frac{\partial^{2}\Psi }{\partial y^{2}}
+ mg\left (y\Psi
-\frac{i\theta}{2}\frac{\partial\Psi}{\partial x}\right)\ +
 \omega J\Theta(y-y_{0})\star  \Psi= E\Psi \,,
\ee
where we have used the fact that the operator $\,H_{c}=\omega J\,$ commutes with the $\,\star\,$-product.
Further, by making use of the 
{\it ansatz} $\,\Psi(x,y,q)=e^{ikx}\,\Phi(y,q)\,$ into (\ref{Schroedinger_equation_2})
we can write
\be \label{Schroedinger_equation_simple}
-\frac{\hbar ^{2}}{2m}\frac{\partial^{2}\Psi }{\partial y^{2}}
+mgy\,\Psi+ \omega J\Theta(y-y_{0})\star \Psi= \left(E-\frac{\hbar^2
 k^2}{2m}-\frac{mgk \theta}{2}\right)\Psi \,.
\ee

It remains to calculate the product $\,\Theta(y-y_0)\star  \Psi\,$. Since $\,\Theta(y-y_0)\,$ 
is not a differentiable function, we cannot directly apply the definition 
of the Moyal product to this case. 
In order to overcome this problem, we have to interpret the term $\,\Theta(y-y_0)\star  \Psi\,$
in the sense of generalized functions (distributions). We start by recalling that the Heaviside
function $\,\Theta(y-y_0)\,$ can be used to define the Heaviside distribution,  
 $\,\Theta : C^{\infty}_{0}(\mathbb{R}) \to \mathbb{C}\,$, given by
\be
\Theta(\phi)=\int_{\mathbb{R}}dy\,\Theta(y-y_{0})\,\phi(y)\,.
\label{distr-theta}
\ee
We say that the function $\,\Theta(y-y_0)\,$ \emph{represents} the 
regular distribution $\,\Theta\,$. 
Even though the Heaviside function is not differentiable,
the corresponding distribution $\,\Theta\,$ is infinitely differentiable. Indeed, recall that given a distribution $\,T\,$, 
its $\,n\,$-th derivative
(in the sense of distributions) 
is defined by 
\be
\left(D^{(n)}T\right)(\phi)=(-1)^{n}\,T\left(\frac{d^{n}\phi}{dy^{n}}\right),
\label{deriv-distr}
\ee
for all $\,\phi\in C^{\infty}_{0}(\mathbb{R})\,$.

Our aim is to construct the $\star$-product between a distribution and an ordinary function.
In order to do this we will use Eq. (\ref{Moyal_product_definition}) as a guide
and apply the definition given by Eq. (\ref{deriv-distr}), so as to make sense of 
$\partial_x\Theta\,$ and $\,\partial_y\Theta$. Since the Heaviside distribution in Eq. (\ref{distr-theta})
acts on functions of the variable $\,y\,$ only (the test functions $\phi(y)$), and bearing in mind 
Eq. (\ref{deriv-distr}),
we consider the following prescription,
\begin{eqnarray}
\frac{\partial}{\partial x}\Theta \longrightarrow &0\,,\qquad
\frac{\partial^{n}}{\partial y^{n}} \Theta \longrightarrow  D^{(n)}\Theta\,.
\end{eqnarray}
Thus, in analogy with Eq. (\ref{Moyal_product_definition}) we can write
\begin{eqnarray}
\Theta\star\Psi
=\Psi(x,y,q)\sum_{n=0}^{\infty}\frac{1}{n!}\left(\frac{k\theta}{2}\right)^{n}D^{(n)}\Theta\,.
\label{first-form}
\end{eqnarray}
The Eq. (\ref{first-form})
shows that the Moyal product $\,\Theta\star \Psi\,$ is equivalent to the product of an ordinary function ($\Psi$)
by another distribution.

In order to achieve a suitable representation for $\,\Theta\star\Psi\,$
we apply Eq. (\ref{first-form}) on a generic test function $\,\phi\in C^{\infty}_{0}(\mathbb{R})\,$. Recaling that
the product $\,uT\,$ between an ordinary function $\,u\,$ and a distribution $\,T\,$ is the distribution defined by
$\,(uT)(\phi)=T(u\phi)\,$, it results that
\begin{eqnarray} \label{def-moyal-distr}
\left(\Theta\star\Psi\right)(\phi) &=& \sum_{n=0}^{\infty}\frac{1}{n!}\left(\frac{-k\theta}{2}\right)^{n}
\int_{\mathbb{R}}dy\,\Theta(y-y_0)\,\frac{d^{n}}{dy^{n}}\left(\Psi(x,y,q)\,\phi(y)\right)\nonumber\\
&=& \int_{\mathbb{R}}dy\,\Theta(y-y_0)\,\Psi\left(x,y-\frac{k\theta}{2},q\right)\,
\phi\left(y-\frac{k\theta}{2}\right),
\end{eqnarray}
what, after the change of variable $\,\xi=y-k\theta/2\,$, leads to
\be
\left(\Theta\star\Psi\right)(\phi) =\int_{\mathbb{R}}d\xi\,\Theta\left(\xi-y_{0}+\frac{k\theta}{2}\right)\,\Psi\left(x,\xi,q\right)\,
\phi\left(\xi\right).
\label{final-form}
\ee

The Eq. (\ref{final-form}) permits us to identify the function $\,(\Theta\star\Psi)(x,y,q)\,$ which represents the
regular distribution $\,\Theta\star\Psi\,$ (in the same sense as the function $\,\Theta(y-y_0)\,$ represents the 
regular distribution $\,\Theta$),
\be
(\Theta\star\Psi)(x,y,q)\equiv\Theta\left(y-y_{\theta}\right)\,\Psi\left(x,y,q\right)\,,
\label{define-funcao}
\ee
where $\,y_{\theta} = y_{0} - k\theta/2\,$.

Now, replacing the term $\,\Theta(y-y_0) \star \Psi(x,y,q)\,$ in Eq. (\ref{Schroedinger_equation_simple}) by Eq. (\ref{define-funcao}), 
we obtain
\be\label{Schr_eq}
-\frac{\hbar ^{2}}{2m}\frac{\partial^{2}\Phi }{\partial y^{2}}
+ mgy\Phi  
+\Theta(y-y_{\theta})\omega J\,\Phi=
\left(E-\frac{\hbar^2 k^2}{2m}-\frac{mgk\theta}{2}\right)\Phi \,,
\ee
where we have made use of the {\it ansatz}  $\Psi (x, y, q) = e^{ikx} \Phi (y, q)\,$.

We remark that the above construction is well-defined only if $k\theta>0\,$. Indeed, 
the Eq. (\ref{def-moyal-distr}) implies that $\,\psi\,$ must be infinitely differentiable in $\,(y_0 , \infty)\,$. But
according to Eq. (\ref{Schr_eq}) the function $\,\frac{\partial^{2}\psi}{\partial y^{2}}\,$ is discontinuous at 
$\,y=y_0-\frac{k\theta}{2}\,$, so that
if $\,k\theta<0\,$ then $\,\psi\,$ is not smooth in $\,(y_{0} , \infty)\,$. 
This result could be associated to a possible time-reversal symmetry breaking in our non-commutative model 
(the effect of a time reversal can be
obtained by the replacement of $\,k\,$ by $\,-k\,$ in the equations of the model, the other parameters kept unchanged). 
We note that at the level of NCQFTs the CPT symmetry is still preserved, but with individual violations of the C, P and T symmetries 
\cite{Chaichian:2002vw}. In any case, further studies are necessary in order to properly generalize the definition given in 
Eq. (\ref{first-form}).

It is now useful to write the wave function $\Phi(y,q)\,$ just like in
 (\ref{ondafinal}), i.e.,
\begin{equation}\label{just_like_ondafinal}
\Phi (y,q)= \frac{1}{\sqrt{N}}\sum_{n=-j}^j \psi^{n}_k
 (y)u_n (q)\,,
\end{equation}
where $u_n (q)\,$ are the basis vectors defined in (\ref{base1}). 
Substituting (\ref{just_like_ondafinal}) into (\ref{Schr_eq}) we find
\begin{equation}
\frac{1}{\sqrt{N}}\sum_{n=-j}^{j}\left\{-\frac{\hbar
 ^{2}}{2m}\frac{d^{2}\psi^{n}_k}{dy^{2}}+ \left[mgy+
 n\hbar\omega\Theta (y-y_{\theta})
-\mathcal{E}_{\theta}\right] \psi^{n}_k (y)\right\}\,u_{n}(q) = 0\,,
\end{equation}
where 
\begin{equation}
\mathcal{E}_{\theta}=E-\frac{\hbar^2 k^2}{2m}-\frac{mgk\theta}{2}\,.
\label{def-E_teta}
\end{equation}
But since the vectors $u_{n}\,$ are linearly independent, we can write a
 differential equation for each $\psi^{n}_k (y)\,$,
\be
-\frac{\hbar ^{2}}{2m}\frac{d^{2}\psi^{n}_k}{dy^{2}}+ \left[mgy+
 n\hbar\omega\Theta (y-y_{\theta})
-\mathcal{E}_{\theta}\right] \psi^{n}_k (y)=0\,.
\ee

There are two different regions to consider, according to the value
 assumed by the Heaviside function. For $\,y < y_{\theta}\,$ we have
$\Theta(y-y_{\theta})=0\,$, so that the Schr\"{o}dinger equation
 reduces to
\be \label{Sch_eq_reduces}
\frac{d^{2}\psi^{n}_k (y)}{dy^{2}} +\frac{2m^{2}g}{\hbar ^{2}}
\left (\frac{\mathcal{E}_{\theta}}{mg}-y\right)
\psi^{n}_k (y)=0\,,
\ee
whose general solution (see Eq.s (\ref{eq-dif1})-(\ref{soluc-y}))
is
\be \label{solucao_I}
\psi_{k}^{n}(y) = A\,Ai\left(\frac{y-b_{\theta}}{a}\right) + B\,
 Bi\left(\frac{y-b_{\theta}}{a}\right) ,
\ee
where $\,a\,$ and $\,b_{\theta}\,$ were defined by Eq.s (\ref{definicoes}) and (\ref{define-b}), respectively.

For $y\geq y_{\theta}\,$ we have $\,\Theta(y-y_{\theta}) = 1\,$\,, so
 that the Schr\"{o}dinger equation reads
\be \label{schroed_reads}
\frac{d^{2}\psi^{n}_k (y)}{dy^{2}} +\frac{2m^{2}g}{\hbar ^{2}}
\left (\frac{\mathcal{E}_{\theta}^{n}}{mg}-y\right)
\psi^{n}_k (y)=0\,,
\ee
where 
\begin{equation}
\mathcal{E}_{\theta}^{n}=\mathcal{E}_{\theta}-n\hbar\omega
\end{equation}
and $\,\mathcal{E}_{\theta}\,$ was defined by Eq. (\ref{E_theta}).

Equation (\ref{schroed_reads}) is analogous to Eq. (\ref{Sch_eq_reduces}), the only difference being in the value of the energy.
It follows that its general solution is a linear combination of Airy functions too.
Now, notice that the Airy function $Bi\,$ is physically unacceptable in the positive semi-axis,
since it diverges for large
 positive values of its argument. Taking this into account
we write the general solution of the Schr\"{o}dinger equation for
 $y\geq y_{\theta}\,$ as
\be \label{solucao_II}
\psi^{n}_k (y) = C\,Ai\left(\frac{y-b_{\theta}^{n}}{a}\right),
\ee
where 
\be \label{relacao_energia}
b_{\theta}^{n}=\frac{\mathcal{E}_{\theta}^{n}}{mg} = b_{\theta} -
 \frac{n\hbar\omega}{mg}\,.
\ee

The complex constants in (\ref{solucao_I}) and (\ref{solucao_II}) 
are determined by the matching conditions at $y=y_{\theta}\,$. 
By setting $\xi=(y-b_{\theta})/a\,$, we see that 
\be
y=y_{\theta}\,\Longrightarrow\,\xi=\frac{y_{\theta}-b_{\theta}}{a}=\frac{y_{0}-b}{a}\,.
\ee 
In what follows we denote this particular value of $\xi\,$
by the symbol $\xi_0\,$. Thus,
\begin{eqnarray}\label{linear}
A\,Ai\left(\xi_{0}\right)+\,B\,Bi\left(\xi_{0}\right)&=&
C\,Ai\left(\xi_{0}+\frac{n\hbar\omega}{mga}\right),\nonumber\\
A\,\frac{dAi}{d\xi}\left(\xi_0 \right)+\,B\,\frac{dBi}{d\xi}\left(\xi_0 \right)&=&C\,\frac{dAi}{d\xi}
\left(\xi_0 +\frac{n\hbar\omega}{mga}\right)\,.
\end{eqnarray}

By solving the above system we find
the values of $\frac{B}{A}\,$ and $\frac{C}{A}\,$, 
\be \label{matching_conditions_result}
\frac{B}{A}=\frac{C}{A}\,\frac{Ai\left(\xi_0 +\frac{n\hbar\omega}{mga}\right)}{Bi\left(\xi_0 \right)}-
\frac{Ai\left(\xi_0 \right)}{Bi\left(\xi_0 \right)}\,,
\ee

\be \label{matching_II}
\frac{C}{A}=\frac{1}{\pi}\,\frac{1}
{Ai\left(\xi_0 +\frac{n\hbar\omega}{mga}\right)\,\frac{dBi}{d\xi}\left(\xi_0 \right)-
\frac{dAi}{d\xi}\left(\xi_0 +\frac{n\hbar\omega}{mga}\right)\,Bi\left(\xi_0 \right)}\,\,.
\ee
The constant $A\,$ remains undetermined. We will choose its value later.

At this point it is useful to put together the results just
 obtained and write the complete solution 
of (\ref{Schroedinger_equation_2}). At first we
recall that $\Psi(x,y,q)=e^{ikx}\Phi(y,q)\,$. Besides that, $\Phi\,$ was
 decomposed according to (\ref{just_like_ondafinal}). We thus write
\be \label{auto_estado_completo}
\Psi_{E,k}(x,y,q)=\mathcal{C}_{E,k}\,\frac{e^{ikx}}{\sqrt{N}}\sum_{n=-j}^j
 \,\psi_{k}^{n}(y)\,u_{n}(q) \,,
\ee
where 
\begin{equation}\label{large}
\psi_{k}^{n}(y) = \left\{
\begin{array}{cccc}
A\,Ai\,\left(\frac{y-b_{\theta}}{a}\right)+B\,Bi\left(\frac{y-b_{\theta}}{a}\right)
 &, \qquad\mbox{if} & y\leq y_{\theta}&,\\
C\,Ai\left(\frac{y-b_{\theta}^{n}}{a}\right) & ,\qquad\mbox{if} &y\geq
 y_{\theta}&,
\end{array}
\right.
\end{equation}
with $B/A\,$ and $C/A\,$ given by (\ref{matching_conditions_result}) and 
(\ref{matching_II}).$\,\,\mathcal{C}_{E,k}\,$ is a normalization constant.

We interpret the eigenstate $\Psi\,$ as the state of a particle with definite energy, traveling within a region of 
uniform gravitational field and interacting with a quantum clock. In
 order to determine the time of flight, we follow the ideas presented in
Sec. \ref{time_section} and suppose
that the interaction between the particle and the clock barrier can be
 treated as a small perturbation. 
At this point it is interesting to compare the present situation with that
 of the Sec. \ref{time_section}. There, the scattered (transmitted) wave function
  could be written as a product of a free particle wave function by a clock state 
carrying the information about the time of flight. 
In the gravitational case the situation is analogous, since that
in the limit of small perturbations the only effect of the interaction
particle-clock is to make the clock run and record the time of flight. 
In the following, we will show that in the ``far future''
the scattered (reflected) wave function can be written as a product of
 a wave function 
of a particle under the action of a purely gravitational potential (without the clock) by a
 clock state showing the time of flight.

We now turn our attention to the
 reflected wave function in the ``far future''. In the stationary framework
that means that we must choose $\xi<0\,$ and $|\xi|\,$ large.  
This fact allows us to make use of the asymptotic form of
the Airy functions in (\ref{large}). If $\,\xi<0\,$ and 
 $\,|\xi|\,$ is sufficiently large we have (see, e.g., \cite{Olver:1974})
\be \label{assint_Ai}
Ai (\xi)\sim\frac{1}{\sqrt{\pi}}|\xi|^{-1/4}\cos\left (
\frac{2}{3}|\xi|^{3/2}-\frac{\pi}{4}\right )  ,
\ee

\be \label{assint_Bi}
Bi (\xi)\sim\frac{1}{\sqrt{\pi}}|\xi|^{-1/4}\sin\left (
\frac{2}{3}|\xi|^{3/2}-\frac{\pi}{4}\right )  .
\ee

The trigonometric functions can be written in terms of complex
 exponentials, and the reflected wave function comes from the exponentials with 
negative exponents. Using (\ref{assint_Ai}) and (\ref{assint_Bi}) into
 (\ref{solucao_I}) we obtain the following expression for the
 reflected wave function
\be \label{assymptotic}
 \psi_{k , ref}^{n}(y) \sim \,A
 \,\frac{1}{2\,\sqrt{\pi}}|\xi|^{-1/4}\,\left(1+i\,\frac{B}{A}\right)\,e^{ -i\left (
\frac{2}{3}|\xi|^{3/2}-\frac{\pi}{4}\right )}\,.
\ee

Analogously, the incident wave function is given by 
\be \label{incident}
\psi_{k , inc}^{n}(y) \sim
 A\,\frac{1}{2\,\sqrt{\pi}}|\xi|^{-1/4}\,\left(1-i\,\frac{B}{A}\right)\,e^{ i\left (
\frac{2}{3}|\xi|^{3/2}-\frac{\pi}{4}\right )}\,.
\ee

We now choose $A\,$ in such a way that $\,A=1+i\,B\,$. With this choice the 
function $\psi_{k}^{n}(y)\,$ reduces to $Ai(\frac{y-b}{a})\,$ in the limit of a negligible clock 
barrier. As a consequence, the reflected and the incident waves read
\be \label{new_reflected}
\psi_{k , ref}^{n}(y) \sim
 \,\left(\frac{1+i\frac{B}{A}}{1-i\frac{B}{A}}\right)\,\frac{1}{2\,\sqrt{\pi}}|\xi|^{-1/4}\,e^{ -i\left (
\frac{2}{3}|\xi|^{3/2}-\frac{\pi}{4}\right )}\,,
\ee

\be \label{new_incident}
\psi_{k , inc}^{n}(y) \sim
 \frac{1}{2\,\sqrt{\pi}}|\xi|^{-1/4}\,e^{ i\left (
\frac{2}{3}|\xi|^{3/2}-\frac{\pi}{4}\right )}\,.
\ee

Writing the complex constant
 $R=(1+iB/A)/(1-iB/A)$ in the exponential form, i.e., $R =
 |R|\,e^{i\,\varphi _{R}} =\,e^{i\,\varphi_{R}}\,$,
we get
\begin{eqnarray}\label{final_reflected}
\psi_{k , ref}^{n}(y) \sim
 \,\frac{1}{2\,\sqrt{\pi}}|\xi|^{-1/4}\,e^{ -i\left (
\frac{2}{3}|\xi|^{3/2}-\frac{\pi}{4} - \varphi_{R}\right )}\,,
\end{eqnarray}
where the phase shift $\,\varphi_{R}\,$ can be expressed as 
\be \label{phase_formula}
\,\varphi_{R}=\arctan
 \left(\frac{2\frac{B}{A}}{1-\frac{B^2}{A^2}}\right).
\ee

The phase shift $\varphi_{R}$ is a function of the energy of the
clock barrier ($E_{cl}=n\hbar\omega$).
In particular, in the absence of the barrier, we have $\,\varphi_{R}(0) = 0\,$, because
if the potential is purely gravitational, the stationary state
 is the Airy function $Ai(\xi)$, as showed in Sec. \ref{noncomm_quant_well}. For a very low energy barrier, i.e., for 
$\,n\hbar\omega<<E\,$, we can expand $\varphi_{R}$ in a power series of $\,n\hbar\omega\,$ and
 restrict it to its first nonvanishing term, the first order correction to the phase shift. This term will give rise to a 
shift in the clock initial state, which will allow us to identify the time of flight of the particle. 

We already know the zeroth order term in the power series expansion
 ($\varphi_{R}(0)=0$). Let us now calculate the first order correction, denoted by $\,\varphi_{R}^{(1)}(n\hbar\omega)\,$.
According to (\ref{phase_formula}), we have
\begin{eqnarray} \label{series_clock}
\varphi_{R}^{(1)}(n\hbar\omega) &=&  \left.\frac{\partial}{\partial E_{cl}}
\arctan\left(\frac{2\frac{B}{A}}{1-\frac{B^2}{A^2}}\right)\right|_{E_{cl}=0}\,\cdot\,n\hbar\omega
\nonumber\\
&=& 2\,\frac{\partial}{\partial
E_{cl}}\left.\left(\frac{B}{A}\right)\right|_{E_{cl}=0}\,\cdot\,n\hbar\omega \,,
\end{eqnarray}
where we have used the fact that $\,\left(B/A\right)(0)=0\,$ (see Eq. (\ref{matching_conditions_result})).

According to Eq. (\ref{matching_conditions_result}) we have
\begin{eqnarray} \label{almost_finished}
\left.\frac{\partial}{\partial E_{cl}}
\left(\frac{B}{A}\right)\right|_{E_{cl}=0} &=& 
\left.\frac{1}{Bi\left(\xi_0 \right)}\,\frac{\partial}{\partial
 E_{cl}} 
\left[
\frac{C}{A}\,Ai\left(\xi_0 +\frac{E_{cl}}{mga}\right)\right]\right|_{E_{cl}=0}\nonumber\\
&=&
\frac{\pi}{mga}\left\{\xi_0\,\left[Ai(\xi_0)\right]^2-
\left[\frac{dAi}{d\xi}(\xi_0)\right]^2\right\}.
\end{eqnarray}

Since the argument $\,\xi_0\,$ is a negative number of large absolute
 value, we can use the asymptotic form of the functions $\,Ai(\xi)\,$ 
and $\,dAi/d\xi\,$. We already know the former (see Eq. (\ref{assint_Ai})). The later can be found
in \cite{Olver:1974}. It reads (up to its leading term)
\be \label{derivada_assintotica}
\frac{dAi}{d\xi}\sim\frac{1}{\sqrt{\pi}}|\xi|^{1/4}\sin\left (
\frac{2}{3}|\xi|^{3/2}-\frac{\pi}{4}\right ).
\ee

Using (\ref{assint_Ai}) and (\ref{derivada_assintotica}) into
 (\ref{almost_finished}) and recalling that $\,\xi_0 < 0\,$ we find
\be
\frac{\partial}{\partial E_{cl}}
\left.\left(\frac{B}{A}\right)\right|_{E_{cl}=0} =
-\frac{1}{\hbar}\sqrt{2\left(\frac{b-y_0}{g}\right)}\,\,\,, 
\ee
so that 
(\ref{series_clock}) is simply given by 
\be \label{resultado_yes}
\varphi_{R}^{(1)}(n\hbar\omega) = -2n\omega\sqrt{\frac{2\left(b-y_{0}\right)}{g}}  \,\,\,.
\ee

We now write $\,T\equiv 2\sqrt{\frac{2\left(b-y_{0}\right)}{g}}\,$ and substitute (\ref{resultado_yes}) into (\ref{final_reflected}) 
 to get the explicit expression of $\,\psi_{k , ref}^{n}(y)\,$,
\be
\psi_{k , ref}^{n}(y) \sim
 \,\frac{1}{2\,\sqrt{\pi}}|\xi|^{-1/4}\,e^{ -i\left (
\frac{2}{3}|\xi|^{3/2}-\frac{\pi}{4}\right)}\,e^{-in\omega T}\,.
\ee

The reflected part of the eigenstate $\Psi(x,y,q)$ (\,see Eq. (\ref{auto_estado_completo})) reads
\begin{eqnarray} \label{resultado_tempo_voo}
\Psi_{E,k,ref}(x,y,q)
&=&\mathcal{C}_{E,k}\,\frac{e^{ikx} }{2\sqrt{\pi}}\,|\xi|^{-1/4}\,e^{ -i\left (
\frac{2}{3}|\xi|^{3/2}-\frac{\pi}{4}\right)}\,
\sum_{n=-j}^j   \frac{e^{in(q-\omega T)}}{\sqrt{2\pi N}} \\\nonumber
&=&\mathcal{C}_{E,k}\,\frac{e^{ikx}}{2\sqrt{\pi}}\,|\xi|^{-1/4}\,e^{ -i\left (
\frac{2}{3}|\xi|^{3/2}-\frac{\pi}{4}\right)}\,  v_{0}(q-\omega T)\,.
\end{eqnarray}

According to Sec. \ref{time_section} (see Eq. (\ref{tempo2}) and the discussion below that equation)  we 
recognize the shifted clock state $v_{0}(q-\omega T)$ as one of the factors in the reflected wave function, as expected.
We can thus finally conclude that
the time of flight of a quantum particle in a uniform gravitational field, in the context of NCQM, is given by
\be \label{time_theta}
T = 2\sqrt{\frac{2\left(b-y_{0}\right)}{g}}\,\,\,.
\ee
In order to understand the physical meaning of the above result, we remark that Eq. (\ref{time_theta}) is equal to the corresponding 
expression of the ordinary quantum mechanics \cite{Davies:2006}. More than this, it is equal to the classical result, 
thus showing that the Weak Equivalence Principle extends 
to the case of quantum mechanics with space-space non-commutativity of the canonical type.

To finish, it is important to clarify a difference between the use we did of Peres approach 
\cite{Peres:1980} in computing the time of flight of a particle in a uniform gravitational field (in NCQM) 
and the use of it by Davies (in ordinary quantum mechanics) \cite{Davies:2006}.  

First of all, it is important to bear in mind that the central idea 
of Peres is to couple a particle to a quantum system (that plays the 
role of a quantum rotor), which will work as a quantum clock. When this  
coupling is considered small, it is such that the effect of the particle-clock 
interaction is essentially only to cause a displacement of the (initial) position 
of the clock pointer, i.e., to change the initial state of the clock. Thus, 
at the end of the interaction, the clock will record the time of flight 
of the particle, which is simply the time interval during which 
the particle interacted with the clock-rotor. 

In order that one can ``watch'' the final position of the clock pointer, and thus 
get the time of flight of the particle, it is necessary that a measurement 
be realized. This, by its turn, must be done in the ``distant future'', when 
the particle-clock interaction can be considered totally negligible, which is 
one of the key points of Peres approach. By analysing the scattered wave function, 
we have thus shown that the time of flight is codified in its phase shift (see Eq.s (\ref{resultado_yes})-(\ref{time_theta})). This result is totally in accordance with 
the spirit of that proved by Peres in the case of a free particle \cite{Peres:1980}.

In the approach by Davies for computing the 
time of flight of a particle in a uniform gravitational field, the Hamiltonian 
of the system does not contain the coupling term between the particle and the 
clock. In this case, the solution of the scattering problem do not ``see'' 
the clock, but only the gravitational barrier. As such, what Davies determined  
was not the phase shift of the wave function due to the interaction between the 
particle (under the action of gravity) and the clock barrier. In fact, what 
he has done was to calculate the derivative (with respect to the total 
energy) of the phase difference induced by the action solely
of gravity on the particle wave function. This derivative was calculated at 
the point where the clock detector should have been placed (it should be 
noted that this derivative is position dependent). Nevertheless, 
it follows that the final result for the time of flight obtained by Davies 
turns out to be the same one that would be obtained by Peres approach, described 
above. Hence, in Davies approach the clock is introduced ad hoc for 
the computation of the time of flight, through the prescription that the 
derivative of the phase difference of the particle wave function has to 
be calculated at the point which corresponds to the position where the clock 
detector should be placed. We remark that in Peres approach, which is based 
on the calculation of the phase shift of the wave function, there is no 
need of dealing with position dependent phases.  

\section{Concluding remarks} \label{conclusao}

We have studied the motion of a particle under the action of a uniform gravitational field in 
NCQM. Assuming non-commutativity only 
on configuration space, we have \emph{exactly} solved 
the non-commutative Schr\"odinger equation and determined the energy eingenvalues after we carefully studied 
the self-adjointeness of the operator involved as well as determined its self-adjoint extensions. We thus concluded 
that the usual boundary condition associated to the reflecting mirror at the botton of the gravitational well 
is among those permitted by the theory of self-adjoint extensions when applied to the original operator we started from. As in ordinary
quantum mechanics, in the non-commutative case the solution is given
by the Airy function and the energy eigenvalues are expressed in
terms of the zeroes of the Airy function $\,Ai\,$. The obtention of the energy
spectrum is specially important, since from data of 
the gravitational quantum well experiment with freely falling
neutrons (the GRANIT experiment) \cite{Nesvizhevsky:2002,Nesvizhevsky:2005ss} 
we could then set an upper-bound on the value of the spatial non-commutative
parameter, $\,\theta\,$. This experimental result can be improved in the future, when
more accurate experimental data will be available \cite{Nesvizhevsky:2005ss}.
We note that the works that considered 
the non-commutative gravitational quantum well have not established  
an upper-bound on $\,\theta\,$, but rather on the 
momentum-momentum non-commutativity parameter \cite{Bertolami:2005jw}-\cite{Buisseret:2007qd} 
or on the time-space component of the non-commutative matrix, 
$\,\theta_{01}\,$ \cite{Saha:2007}.

Another issue we have addressed in this paper was related to the question of 
the (Weak) Equivalence Principle, its validity in NCQM. We were interested in investigating the status of the 
Equivalence Principle through a kind 
of (balistic) Galileo experiment, associated to delocalized, energy eigenstates. 
For that, we have used a quantum clock model, due to Peres \cite{Peres:1980}, 
in order that the time of flight of 
the particle could be measured. We remark that although we were inspired 
by Davies \cite{Davies:2006} in asking for a possible violation of the 
Equivalence Principle by studying the time of flight of a particle subjected 
to a uniform gravitational field, we have performed a conventional stationary state analysis 
of a scattering problem, instead of just naively applying the formula used by Davies 
in the ordinary case \cite{Davies:2006}. Instead of that, we closely 
followed the original approach due to Peres \cite{Peres:1980} of applying a quantum clock 
in the measurement of the time of flight of a quantum particle (see also \cite{Leavens:1993}). It resulted that 
the time of flight is the same as in quantum mechanics, which in turn 
is identical to the classical result, when the measurement is made far from 
the turning point. 
This result can be interpreted as an extension of
the Equivalence Principle to the realm of NCQM.

In order to address until to what extent the (Weak) Equivalence Principle 
holds in NCQM (and analogously in ordinary quantum mechanics) further 
studies are important, such as, for instance, the investigation  
of how matter and anti-matter, in a non-commutative background, 
behave under the action of a gravitational field. Although it has been 
shown that in NCQFTs the CPT symmetry is still 
preserved \cite{Chaichian:2002vw}, but with individual violations of 
the C, P and T symmetries, such a symmetry might be broken at a more 
fundamental level, when not only the matter fields are 
quantized but also the gravitational field itself. In fact, it has been suggested 
that quantum gravity effects might lead to violations of the 
CPT symmetry \cite{Hawking:1976ra}. Finally, we recall that as the 
non-commutativity of space-time might be a signature of quantum gravity, experiments 
involving an interplay between gravity and quantum mechanics are welcome, since even 
at low-energies, but with enough level of accuracy, they might display information 
about the physics whose origin is in a more fundamental level (see \cite{AmelinoCamelia:1999gg}). 

It would be interesting to experimentally investigate the Equivalence Principle 
by measuring the time of flight of quantum particles using a quantum clock. Of course  
that one has to deal with the intrinsic uncertainty of using a quantum clock, as even at low 
energy (weak coupling between the clock and the system of interest), a good 
measurement of the time of flight is subjected to a lower limit on the time 
resolution of the clock \cite{Peres:1980}. In this direction, we quote 
the work of Alonso et al.\cite{Alonso:2003}, which has shown, in the 
case of a free quantum particle, that it is possible to gain an improvement in 
the measurement of time of flight if the quantum clock does not work continuously 
but rather by means of pulsed couplings. 

\bigskip
\noindent {\bf Acknowledgments}\par
\vspace{0.2cm}
\noindent
The authors would like to dedicate this work \textit{in memoriam} to 
\textit{Ivens Carneiro}, a great friend and admirable physicist, who 
prematurely passed away. The authors would like to thank the referee 
for his/her comments which led to improvements in the paper. The authors are grateful to Prof. D. A. T. Vanzella for a helfpul discussion.
K. H. C. Castello-Branco acknowledges Prof. D. A. T. Vanzella, for encouragement,
and also Grupo de F\'isica Te\'orica, do Departamento de F\'isica e Inform\'atica, do Instituto de F\'isica de S\~ao Carlos, for 
hospitality.
A. G. Martins thanks the Universidade de Bras\'ilia for the hospitality
during the preparation of part of this work.

\providecommand{\href}[2]{#2}\begingroup\raggedright\endgroup


\begin{thebibliography}{10}
\bibitem{dense}A subset $\,\mathcal{D}\,$ of $\,\mathcal{H}\,$ is dense (in the topology of $\,\mathcal{H}\,$)
if for every $\,\phi\in \mathcal{H}\,$ there exists a sequence $\,\{\phi_{j}\}\,$, $\,\phi_j\in \mathcal{D}\,$,
such that $\,\phi_j\to\phi\,$ as $\,j\to\infty\,$.
Besides, we have $\,\overline{D(H)}=\mathcal{H}\,$, where
$\,\overline{D(H)}\,$ is the closure of $\,D(H)\,$.
\bibitem{adjointop}Let $\,T\,$ be a densely defined operator and 
$\,D(T^{*})=\lbrace\psi\in\mathcal{H} : \exists \eta\in\mathcal{H} \,\,|\,\,
(T\phi,\psi)=(\phi,\eta)\,,\forall\,\phi\in D(T)\rbrace\,$. We define 
the adjoint $\,T^{*}\,$ of $\,T\,$ as $\,T^{*}\psi=\eta\,$, $\,\forall \psi \in D(T^{*})\,$.
\bibitem{weakder}The $\,n\,$-th weak derivative of the integrable function $\,u\,$ is the integrable function  $\,v\,$ 
such that, for all $\,\phi\in C^{\infty}_{0}(a , b)\,$, we have
$\,\int_{a}^{b} d\xi\, u(\xi)\,\partial_{\xi}^{n}\phi(\xi)=(-1)^{n}\int_{a}^{b} d\xi\, v(\xi)\,\phi(\xi)\,$.
\bibitem{closedop}An operator $\,T\,$ is called closed if and only if its graph, defined as 
$\,\mathcal{G}(T)=\left\lbrace(\phi , \varphi)\in \mathcal{H}\times\mathcal{H} : \varphi=T\phi\right\rbrace$, is closed.
If an operator $\,T\,$ has a closed extension, then it is called closable. 
The closure of a closable operator $\,T\,$ is the operator  $\,\overline{T}\,$ defined by
$\,\mathcal{G}({\overline{T}})=\overline{\mathcal{G}(T)}\,$.
It turns out that $\,\overline{T}\,$ is the smallest closed extension of $\,T\,$.
\end{thebibliography}
\end{document}